\documentclass[preprint,9pt,3p]{elsarticle}
\usepackage{graphicx}
\usepackage{subfigure}
\usepackage{amsmath}
\usepackage{amsfonts}
\usepackage{color}

\biboptions{sort,comma,compress}
\DeclareMathOperator{\sech}{sech}
\DeclareMathOperator{\sn}{sn}
\DeclareMathOperator{\cn}{cn}
\DeclareMathOperator{\dn}{dn}
\begin{document}
\begin{frontmatter}

\title{Rogue wave, interaction solutions to the KMM system}
\author[]{Xin-Wei Jin}
\author[]{Ji Lin$^{\ast}$}
\address{Department of Physics, Zhejiang Normal University, Jinhua, Zhejiang 321004, P.R. China}
\cortext[cor1]{Corresponding author email: linji@zjnu.edu.cn (Ji Lin)}

\begin{abstract}
In this paper, the consistent tanh expansion (CTE) method and the truncated Painlev$\acute{\rm e}$ analysis are applied to the Kraenkel-Manna-Merle (KMM) system, which describes propagation of short wave in ferromagnets. Two series of analytic solutions of the original KMM system (free of damping effect) are obtained via the CTE method. The interaction solutions contain an arbitrary function, which provides a wide variety of choices to acquire new propagation structures. Particularly, the breather soliton, periodic oscillation soliton and multipole instanton are obtained. Furthermore, we obtain some exact solutions of the damped-KMM equation at the first time. On the other hand, a coupled equation containing quadri-linear form and tri-linear form for the original KMM system is obtained by the truncated Painlev$\acute{\rm e}$ analysis, and the rogue wave solution and interaction solutions between rogue wave and multi-soliton for the KMM system are discussed.

\end{abstract}


\end{frontmatter}
\section{Introduction}

Nonlinear partial differential evolution equations play an important role in various physical fields including electromagnetics, fluid mechanics, plasma physics, nonlinear optics and so on \cite{1,2,3,4,5,6}. As a result, in order to predict and explain physical phenomena, exploring exact solutions of these partial differential equations is becoming a primary project. In the past few years, many powerful methods have been developed to search for multiple solutions, such as the Hirota bilinear method \cite{7}, Lie group method \cite{8}, Darboux transformation\cite{9,10}, B\"acklund transformation \cite{11}, the inverse scattering transformation \cite{12}, etc. However, in addition to multiple soliton solutions, it is rather hard to obtain interaction solutions between soliton and other types of nonlinear excitations. Recently, Lou and his cooperators brought forward the consistent tanh expansion (CTE) method \cite{13,14,15}, which can be used to find interaction solutions such as soliton-periodic waves and soliton-cnoidal waves. Based on this effective method, many new interaction solutions for various nonlinear systems have been discussed in detail \cite{16,17,18,19,20,21,22}.

On the other hand, rogue waves are a kind of rational function solutions localized in both space and time, which has attracted an increasingly attention from both mathematics and physics and risen in many fields such as oceanography, Bose-Einstein condensate and nonlinear optics \cite{23,24,25,26,27,28,29,30,31}. Furthermore, it is also worth searching for interaction solutions between rogue waves and multi-solitons in some new systems. This constitutes another purpose of this paper.

In this paper, we focus on the (1+1)-dimensional damped Kraenkel-Manna-Merle (KMM) system \cite{32,33}
\begin{equation}
\begin{split}
&A_{xt}=-BB_{x},\\
&B_{xt}=B A_{x}-sB_{x},
\end{split}
\end{equation}
which is derived from Maxwell's equations supplemented by the Landau-Lifshitz-Gilbert equation and the parameter $s$ stands for the Gilbert damping effect, $x$ and $t$ represent the space-like and time-like variable, respectively. The damped-KMM system (1) was constructed to describe the propagation of electromagnetic short-waves in a saturated ferrite, and the two components of magnetic field can be obtained from observables $A$ and $B$ by relation $H^{y}=-m(1+A_{x}),H^{z}=-mB_{x}$ (where $m$ denotes dimensionless saturation magnetization, we assume $m=1$ in subsequent discussions).

Neglecting the damping effect ($s=0$), the damped-KMM system is reduced to the original KMM system and it can be mapped to the completely integrable sine-Gordon equation. While, the damped-KMM system ($s\ne0$) was proved to be Painlev$\acute{\rm e}$ property non-integrable. To our knowledge, no analytical solution has been obtained for this damped-KMM system. As a matter of fact, all the contributions of the KMM system were built on vanishing the dissipative parameter $s\equiv0$. In Ref.[34], the multi-loop-soliton solutions were obtained via Hirota's bilinearzation, it was found that the interactions depend strongly on the ratio of the amplitudes of the interacting structures. By applying the inverse scattering transformation method and the Wadati-Konno-Ichikawa scheme, the one-, two-, and three-loop and spike-like soliton solutions were constructed in \cite{35}. A class of multiple-valued solitons were investigated by the generalized $G/G'$-expansion method \cite{36}. The loop-like traveling wave solutions and loop-like soliton solutions were obtained via the auxiliary equation method \cite{37,38}. In this work, we restrict our interest to investigate new analytically interaction solutions of the original and damped-KMM system.

The organization of this paper is settled as follows. In Section 2, the CTE method will be applied in the original KMM system ($s$=0). New interaction solutions including breather soliton and multipole instanton wave are obtained. In Section 3, by providing a Painlev$\acute{\rm e}$-B\"acklund transformation, we construct the rogue wave solution, interaction solution between rogue wave and one-soliton, interaction solution between rogue wave and two-soliton for the original KMM system. In Section 4, we present some exact solutions of the damped-KMM system, which has not been obtained in previous work. The last section is devoted to summary and discussion.

\section{Consistent tanh expansion for the original KMM system}
In this section, we aim at seeking for new interaction solutions of the original KMM system. By letting $u=A_{x}$ and $v=iB$, we can transform the original KMM system ($s$=0) into following form
\begin{equation}
\begin{split}
&u_{t}=vv_{x},\\
&v_{xt}=uv.
\end{split}
\end{equation}

For this original KMM system (2), according to the CTE method, the truncated tanh function expansions of functions $u$ and $v$ in (2) have the forms (The order of truncation can be determined by the leading-order analysis)
\begin{equation}
\begin{split}
u=u_{0}+u_{1}\tanh(f)+u_{2}\tanh^{2}(f),\ \ \ \ v=v_{0}+v_{1}\tanh(f),
\end{split}
\end{equation}
where $f$ is an undetermined function of the variables $x$ and $t$, and all of functions $u_{0},u_{1},u_{2},v_{0},v_{1}$ should be determined by vanishing the coefficients of each power of $\tanh(f)$. Then, we obtain
\begin{equation}
\begin{split}
&u_{2}=2f_{x}f_{t},\ \ \ \ v_{1}=2f_{t},\ \ \ \ u_{1}=-2f_{xt},\ \ \ \ v_{0}=-\frac{f_{tt}}{f_{t}},\ \ \ \ u_{0}=-2f_{x}f_{t}+\frac{f_{xtt}}{f_{t}}-\frac{f_{xt}f_{tt}}{f_{t}^{2}},
\end{split}
\end{equation}
and the solution of the original KMM system (2)
\begin{equation}
\begin{split}
&u=-2f_{x}f_{t}+\frac{f_{xtt}}{f_{t}}-\frac{f_{xt}f_{tt}}{f_{t}^{2}}-2f_{xt}\tanh(f)+2f_{x}f_{t}\tanh^{2}(f),\ \ \ \ v=-\frac{f_{tt}}{f_{t}}+2f_{t}\tanh(f),
\end{split}
\end{equation}
where the function $f$ is satisfied the following equation
\begin{equation}
\begin{split}
4f_{xt}f_{t}^{4}-f_{xttt}f_{t}^{2}+f_{xt}f_{t}f_{ttt}+3f_{xtt}f_{t}f_{tt}-3f_{xt}f_{tt}^{2}=0,
\end{split}
\end{equation}
once the solution of Eq.(6) is obtained, then the solution of the original KMM system (2) can be given directly from relation (5).

Apparently, Eq.(6) possesses a trivial solution
\begin{equation}
\begin{split}
f=k_{0}x+\omega_{0} t,
\end{split}
\end{equation}
which lead to the one-soliton solution of the original KMM system (2)
\begin{equation}
\begin{split}
&u=-2k_{0}\omega_{0}\sech^{2}(k_{0}x+\omega_{0} t),\ \ \ \ v=2\omega_{0}\tanh(k_{0}x+\omega_{0} t).
\end{split}
\end{equation}
Furthermore, in order to obtain the interaction solutions of the original KMM system (2), we consider $f$ in the form
\begin{equation}
\begin{split}
f=k_{0}x+\omega_{0} t+F(X),\ \ \ \ X=k_{1}x+\omega_{1}t,
\end{split}
\end{equation}
where $k_{0},\omega_{0},k_{1},\omega_{1}$ are arbitrary constants. Substituting (9) into (6), we have
\begin{equation}
\begin{split}
G_{X}^{2}&=4G^{4}+\left(\frac{8(\omega_{0}-a_{2})}{\omega_{1}}\right)G^{3}+\left(\frac{a_{1}\omega_{1}^{2}}{4}+\frac{4(a_{2}^{2}-4a_{2}\omega_{0}+\omega_{0}^{2})}{\omega_{1}^{2}}\right)G^{2}\\
&+\left(\frac{a_{1}\omega_{0}\omega_{1}}{2}+\frac{8a_{2}\omega_{0}(a_{2}-\omega_{0})}{\omega_{1}^{3}}\right)G+\frac{a_{1}\omega_{0}^{2}}{4}+\frac{4a_{2}^{2}\omega_{0}^{2}}{\omega_{1}^{4}},
\end{split}
\end{equation}
where $G=F_{X}$ and $a_{1}$, $a_{2}$ are free constants. It should be noted that the resulting class of solutions of Eq.(10) will serve to yield a class of interaction solutions to the original KMM system (2) from (5) and (9).

Two special pairs of solutions containing an arbitrary function are generated through solving Eq.(10), which lead to two classes of interaction solutions
\begin{equation}
\begin{split}
&u=-2(k_{0}+P_{1x})\omega_{0}\sech^{2}(k_{0}x+\omega_{0} t+P_{1}),\ \ \ \ v=2\omega_{0}\tanh(k_{0}x+\omega_{0} t+P_{1}).
\end{split}
\end{equation}
and
\begin{equation}
\begin{split}
&u=-2k_{0}(\omega_{0}+P_{2t})\sech^{2}(k_{0}x+\omega_{0} t+P_{2}),\ \ \ \ v=-\frac{P_{2tt}}{\omega_{0}+P_{2t}}+2(\omega_{0}+P_{2t})\tanh(k_{0}x+\omega_{0} t+P_{2}),
\end{split}
\end{equation}
with $P_{1}=P_{1}(x)$ and $P_{2}=P_{2}(t)$.

Hence, a great deal of interaction solutions can be constructed by selecting the arbitrary function as various forms, some meaningful solutions of the original KMM system are given as follows

\textbf{(i) Breather soliton solution}

Taking the arbitrary function as the form $P_{2}=b_{0}+b_{1}\sn(b_{2}t,n)$, we obtain a breather soliton solution
\begin{equation}
\begin{split}
&u=-2k_{0}\left(\omega_{0}+b_{1}b_{2}CD\right)\sech^{2}(\xi),\ \ \ \ v=-\frac{b_{1}b_{2}^{2}S(2n^{2}S-n^{2}-1)}{\omega_{0}+b_{1}b_{2}CD}+2(\omega_{0}+b_{1}b_{2}CD)\tanh(\xi),
\end{split}
\end{equation}
where $\xi=k_{0}x+\omega_{0} t+b_{0}+b_{1}\sn(b_{2}t,n)$ and $S=\sn(b_{2}t,n),C=\cn(b_{2}t,n),D=\dn(b_{2}t,n)$ are the usual Jacobian elliptic functions with the modulus $n$. The structures of two magnetic field components $H^{y}$ and $H^{z}$ are shown in Fig.1 by selecting the arbitrary constants as $k_{0}=2,\omega_{0}=1/2,b_{0}=0,b_{1}=1,b_{2}=1,n=3/10.$ As it is observed, the propagation modes of two magnetic field components are identical except for a phase difference $\pi/2$. The soliton propagating at a constant velocity $\omega_{0}/k_{0}$ towards the positive $x$-axis and its amplitude oscillates periodically with time. In addition, the maximum amplitude appears alternately owing to the phase difference $\pi/2$, which signifies the transfer of energy between two components.
\begin{figure}[htbp]
\subfigure[]{
\begin{minipage}[t]{0.5\linewidth}
\centering
\includegraphics[width=5.3cm]{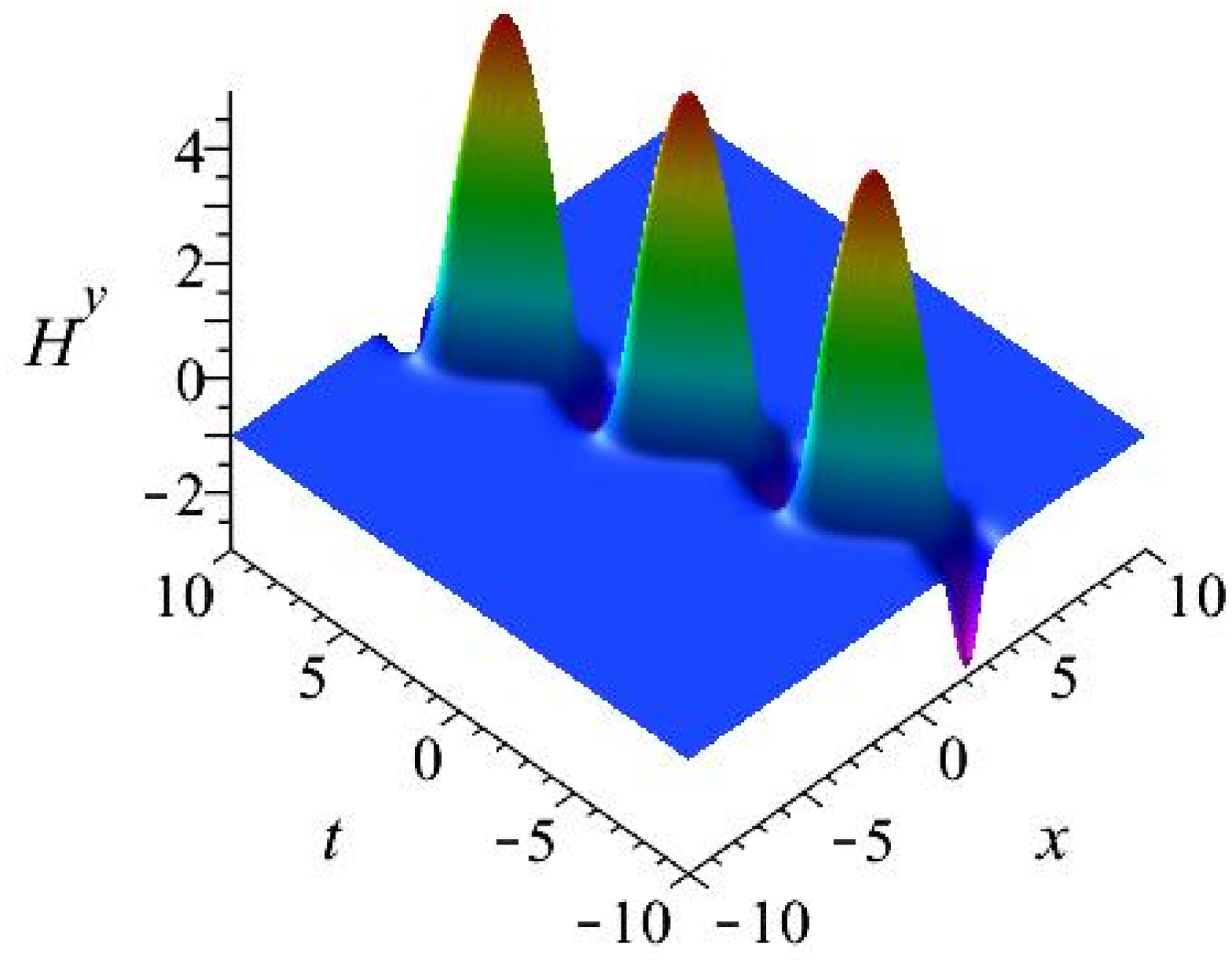}
\end{minipage}%
}
\subfigure[]{
\begin{minipage}[t]{0.5\linewidth}
\centering
\includegraphics[width=5.3cm]{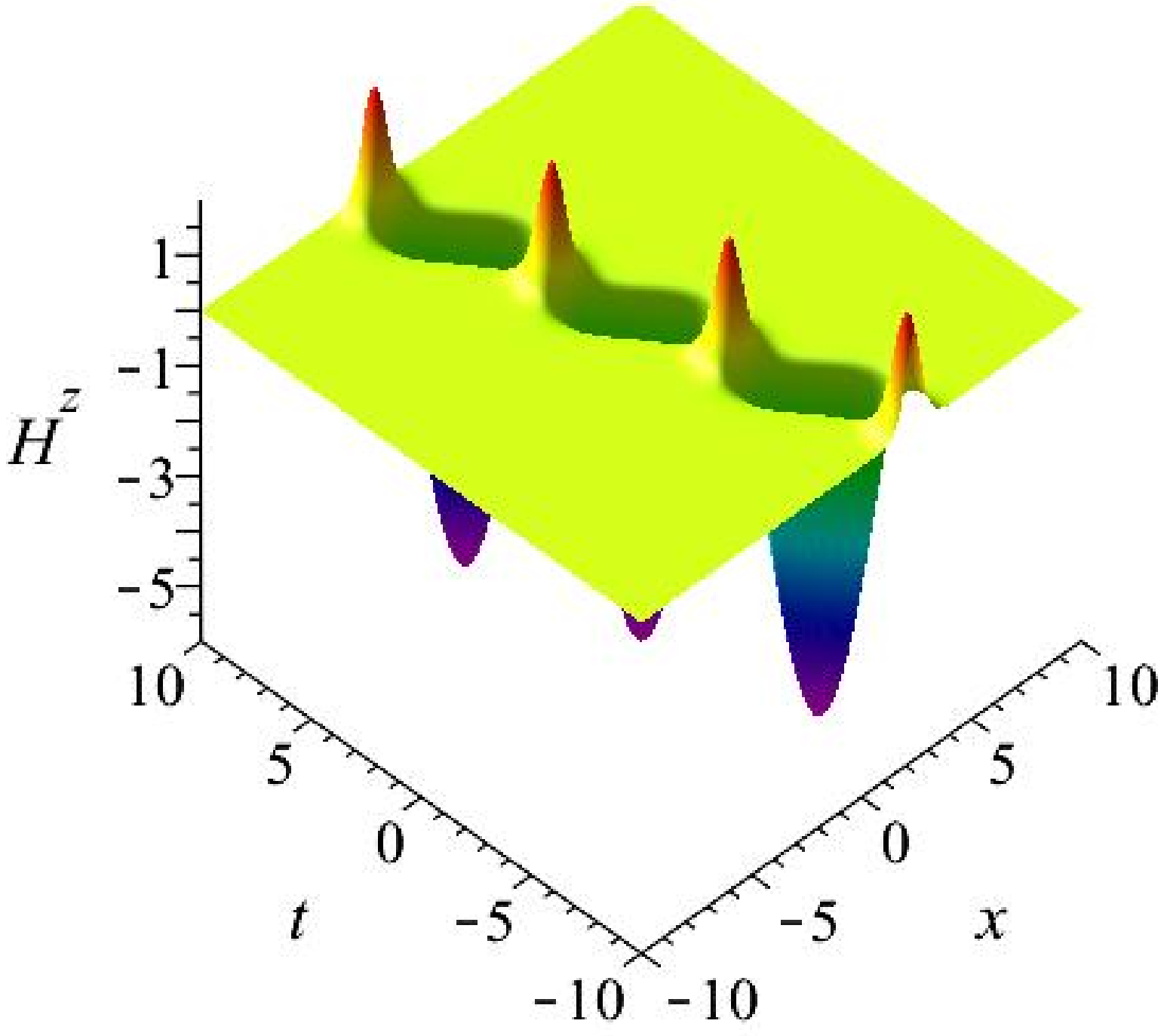}
\end{minipage}
}
\caption{Propagation of breather soliton (13) of (a) $H^{y}$ and (b) $H^{z}$ for $k_{0}=2,\omega_{0}=1/2,b_{0}=0,b_{1}=1,b_{2}=1,n=3/10.$}
\end{figure}
\begin{figure}[htbp]
\subfigure[]{
\begin{minipage}[t]{0.5\linewidth}
\centering
\includegraphics[width=5.3cm]{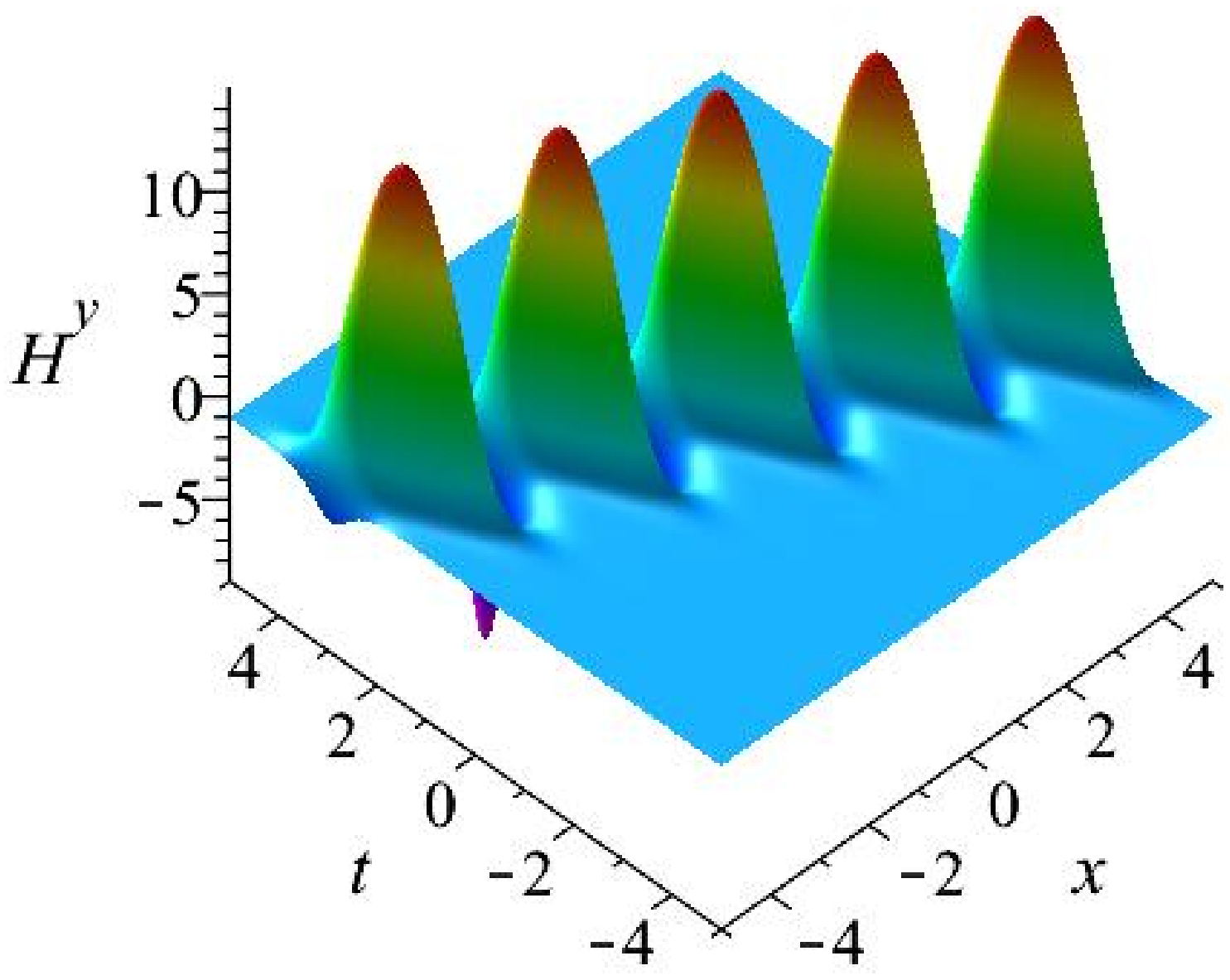}
\end{minipage}%
}
\subfigure[]{
\begin{minipage}[t]{0.5\linewidth}
\centering
\includegraphics[width=5.3cm]{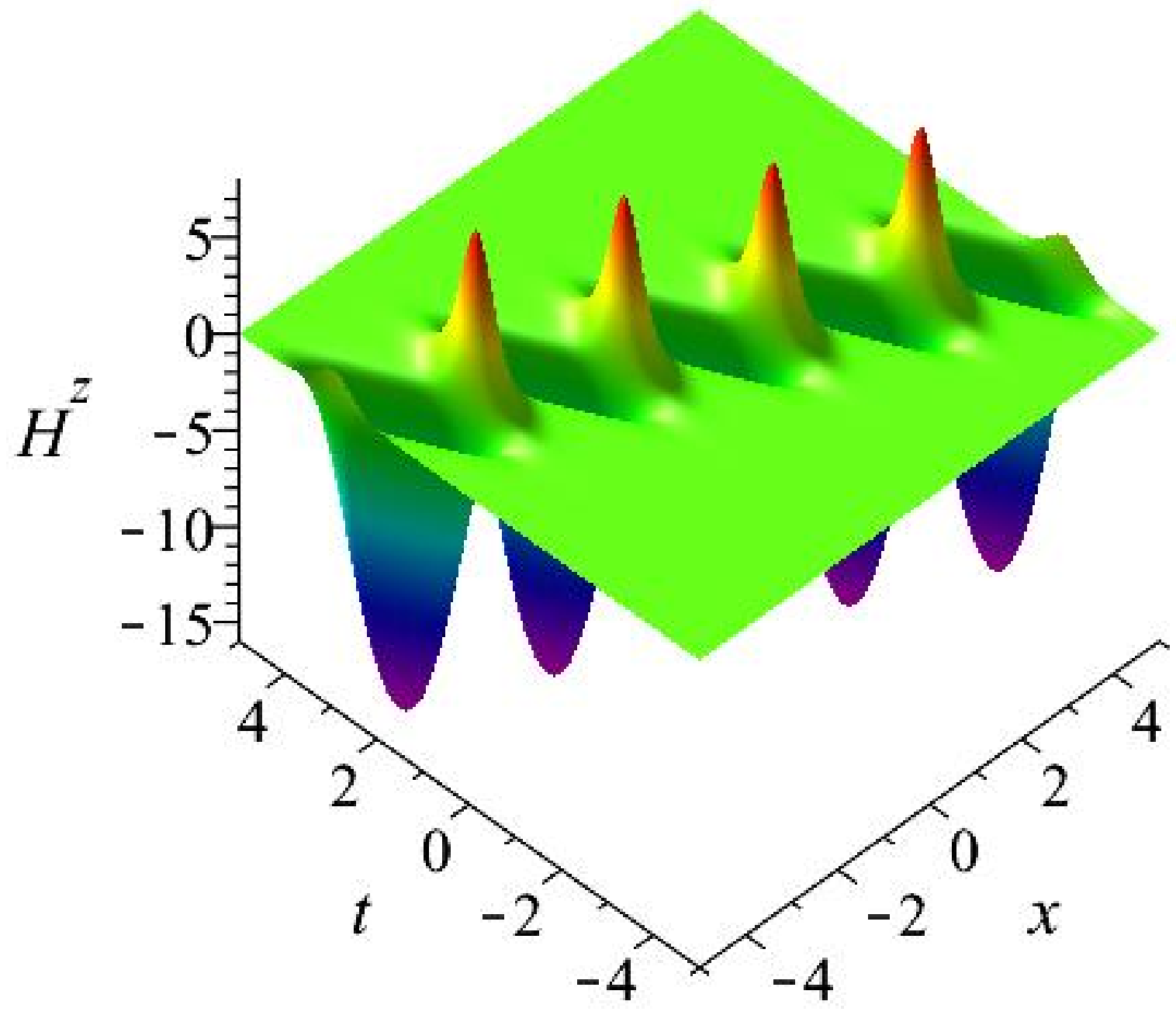}
\end{minipage}
}

\subfigure[]{
\begin{minipage}[t]{0.5\linewidth}
\centering
\includegraphics[width=4.6cm]{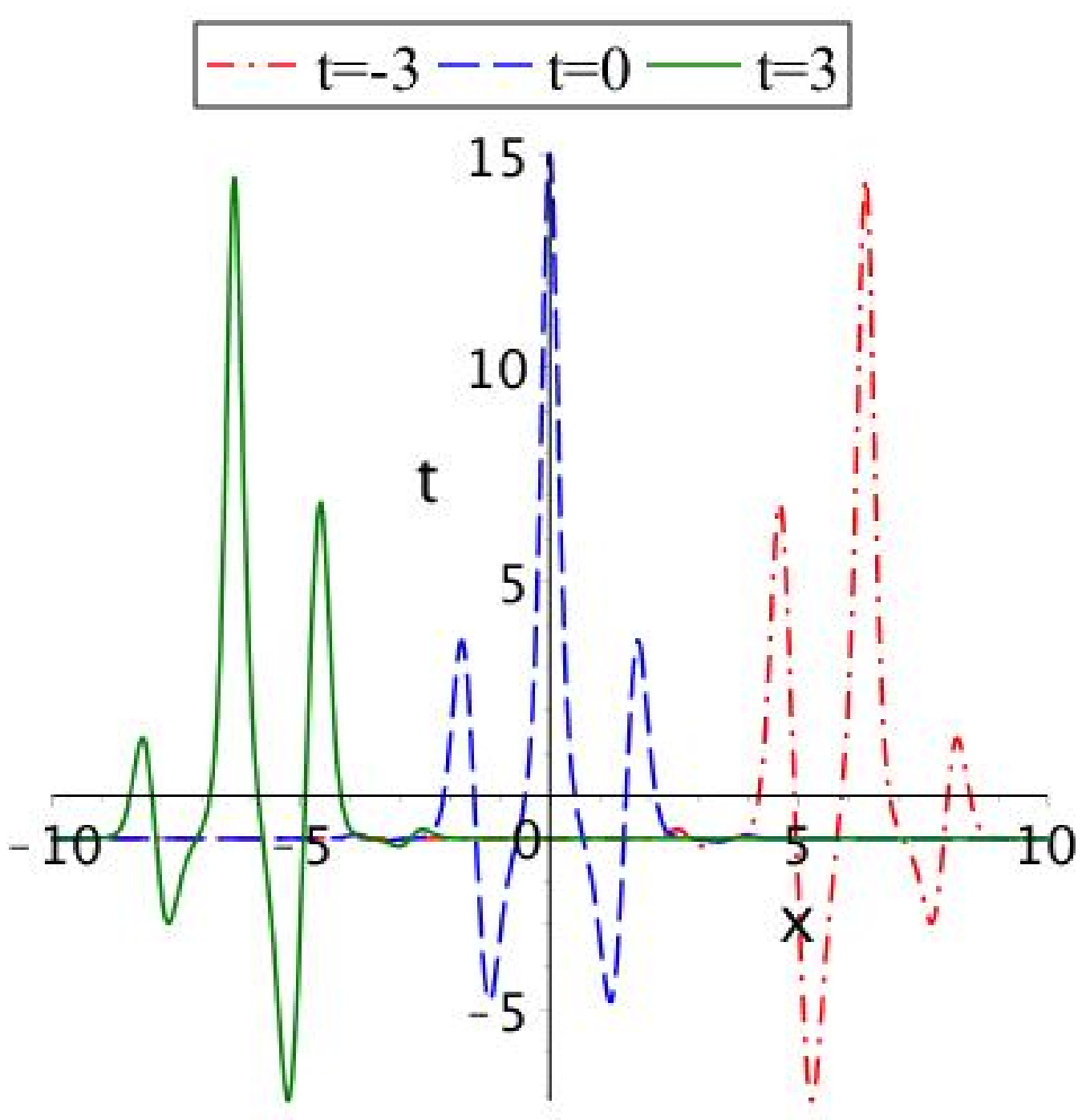}
\end{minipage}%
}
\subfigure[]{
\begin{minipage}[t]{0.5\linewidth}
\centering
\includegraphics[width=4.6cm]{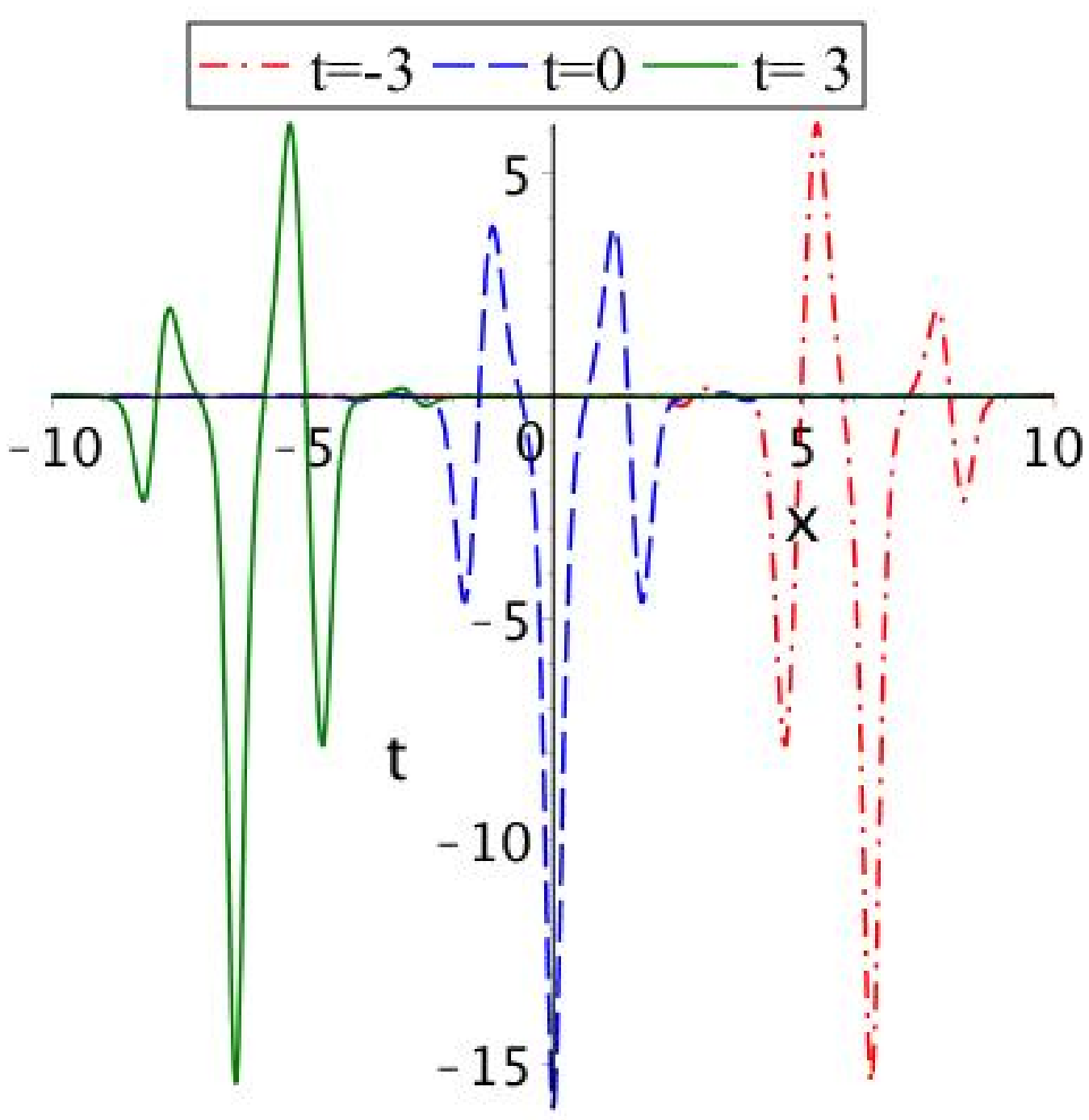}
\end{minipage}
}
\caption{Periodic oscillation soliton (14) with $k_{0}=1,\omega_{0}=2,d_{0}=0,d_{1}=1,d_{2}=3,n=3/10$: (a) $H^{y}$, (b) $H^{z}$, propagation of (c) $H^{y}$ and (d) $H^{z}$ at times $t=-3$, $t=0$, $t=3$.}
\end{figure}

\textbf{(ii) Periodic oscillation soliton solution}

Still paying interests to the interacting waves with periodic function, we consider the arbitrary function as the form $P_{1}=d_{0}+d_{1}\sn(d_{2}x,m)$, the periodic oscillation soliton solution can be obtained
\begin{equation}
\begin{split}
&u=-2\omega_{0}\left(k_{0}+d_{1}d_{2}CD\right)\sech^{2}(\xi),\ \ \ \ v=2\omega_{0}\tanh(\xi),
\end{split}
\end{equation}
where $\xi=k_{0}x+\omega_{0} t+d_{0}+d_{1}\sn(d_{2}x,n)$ and $C=\cn(d_{2}x,n)$,$D=\dn(d_{2}x,n)$ are the usual Jacobian elliptic functions with the modulus $n$. The illustration is made in Fig.2 which presents the spatial structures and the shapes of $H^{y}$ and $H^{z}$ at times $t=-3$, $t=0$, $t=3$, with the arbitrary constants are fixed as $k_{0}=1,\omega_{0}=2,d_{0}=0,d_{1}=1,d_{2}=3,n=3/10$. Unlike the previous propagation mode, this kind of periodic oscillation is reflected in the $x$-direction. Figs.2(c) and 2(d) reveal the fact that this solution is essentially oscillatory propagation of few-cycle pulse. And it is obvious that the two magnetic field components $H^{y}$ and $H^{z}$ can be converted to each other by a constant phase difference $\pi/2$.

\textbf{(iii) Multipole instanton solution}

The multipole instanton solution can be constructed by selecting $P_{2}=\sech(t)\tanh(t)^{3}\cos(2t)-\omega_{0}t$,
\begin{equation}
\begin{split}
u=&2k_{0}\left[4\sech(t)\tanh^{4}(t)\cos(2t)+2\sech(t)\tanh^{3}(t)\sin(2t)-3\sech(t)\tanh^{2}(t)\cos(2t)\right]\sech^{2}(\xi),\\ v=&\frac{20\tanh^{5}(t)\cos(2t)+16\tanh^{4}(t)\sin(2t)-29\tanh^{3}(t)\cos(2t)-12\tanh^{2}(t)\sin(2t)+6\tanh(t)\cos(2t)}{4\tanh^{4}(t)\cos(2t)+2\tanh^{3}(t)\sin(2t)-3\tanh^{2}(t)\cos(2t)}\\
&-\left[8\sech(t)\tanh^{4}(t)\cos(2t)+4\sech(t)\tanh^{3}(t)\sin(2t)-6\sech(t)\tanh^{2}(t)\cos(2t)\right]\tanh(\xi),
\end{split}
\end{equation}
where $\xi=k_{0}x+\sech(t)\tanh(t)^{3}\cos(2t)$. The spatial structures and the contour plots of $H^{y}$, $H^{z}$ are described in Fig.3. As one can analyze the snapshots, when time tends to be negative infinite, the energy of the electromagnetic field is concentrated only on $H^{y}$ as background wave. As time moves forward, both components of the magnetic field $H^{y}$ and $H^{z}$ start to oscillate around the background plane, respectively. During this whole process, the total energy of two magnetic field components remains the same. After the oscillation, both amplitudes of $H^{y}$ and $H^{z}$ return to the initial background.
\begin{figure}[htbp]
\subfigure[]{
\begin{minipage}[t]{0.5\linewidth}
\centering
\includegraphics[width=5.1cm]{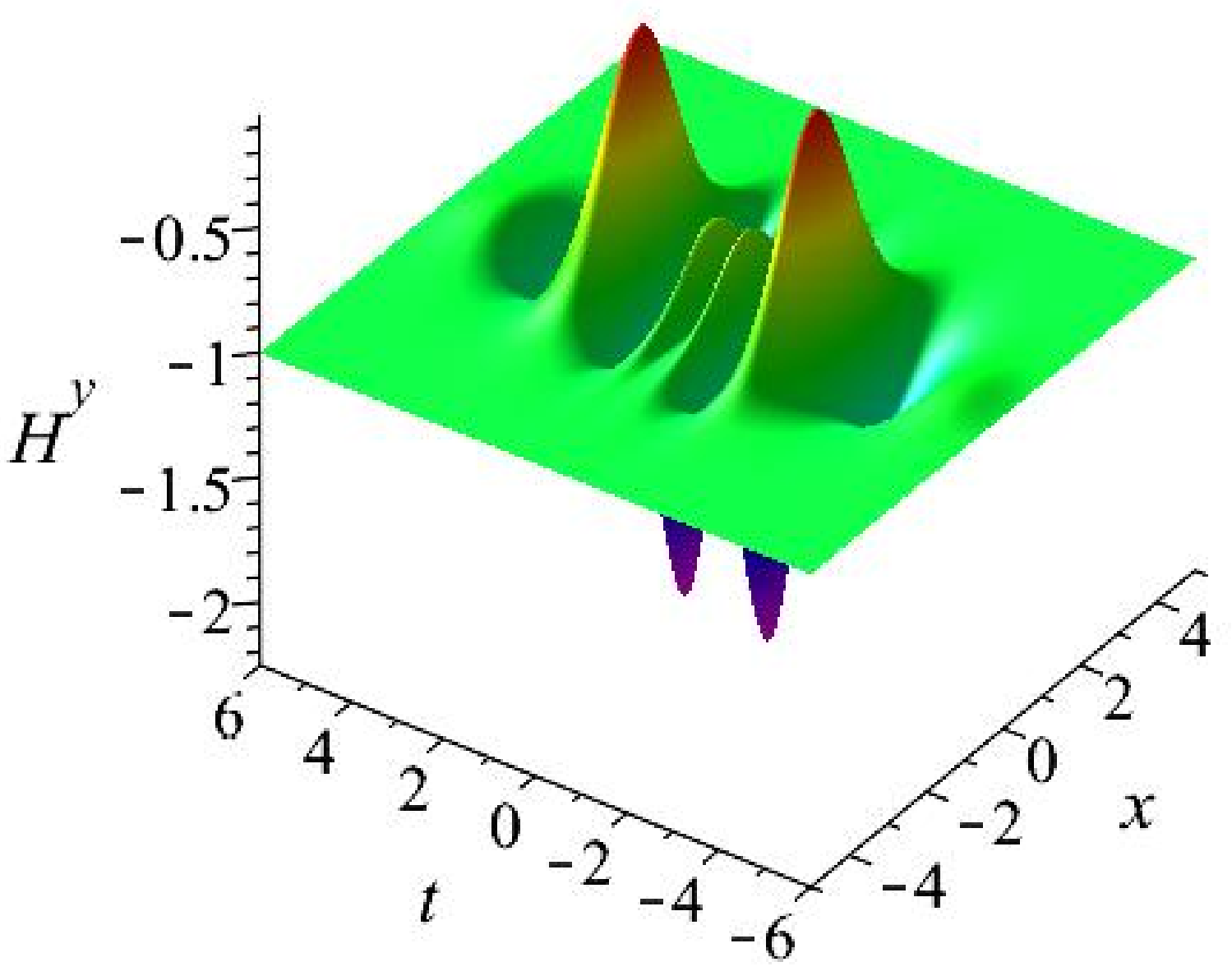}
\end{minipage}%
}
\subfigure[]{
\begin{minipage}[t]{0.5\linewidth}
\centering
\includegraphics[width=5.1cm]{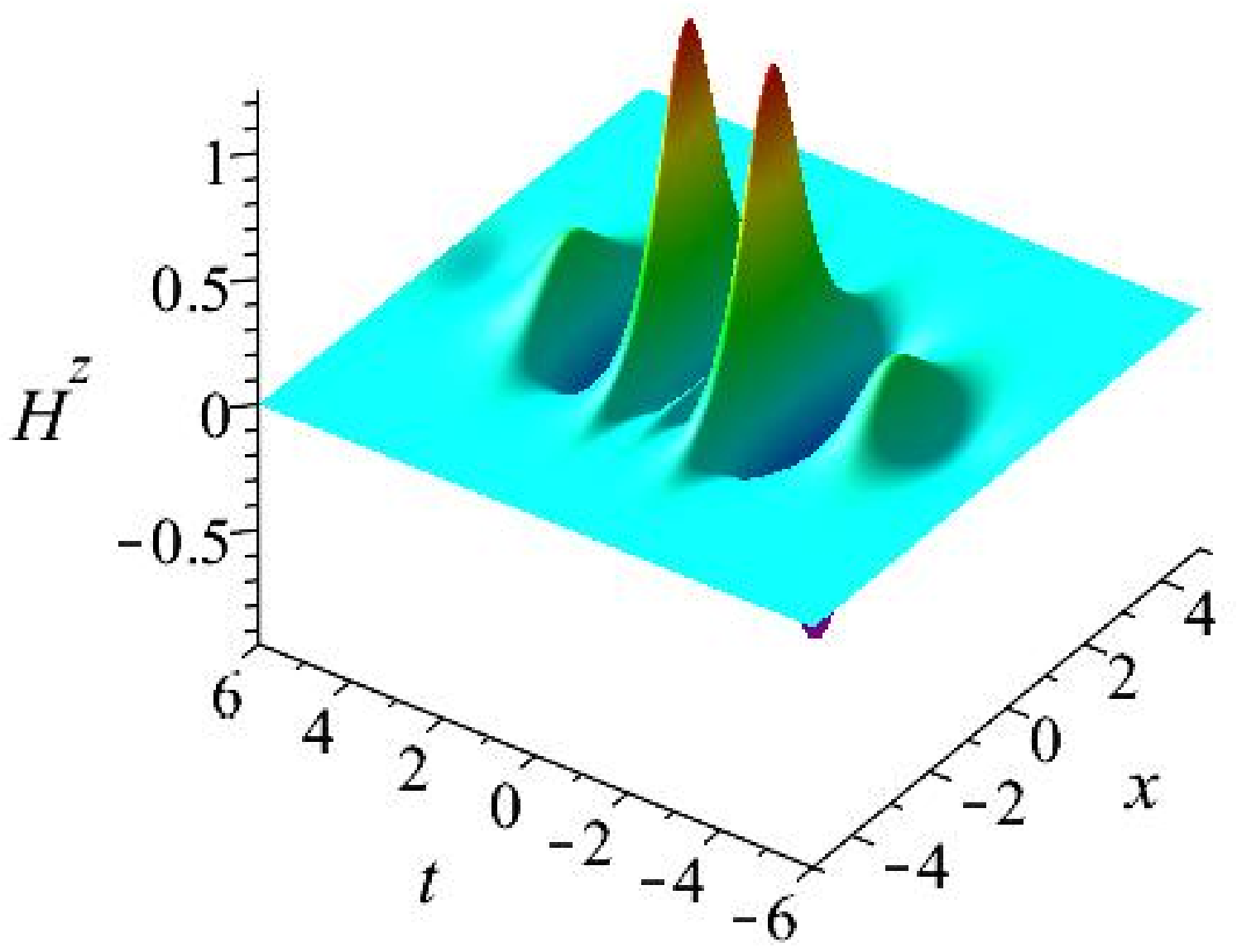}
\end{minipage}
}

\subfigure[]{
\begin{minipage}[t]{0.57\linewidth}
\centering
\includegraphics[width=5.5cm]{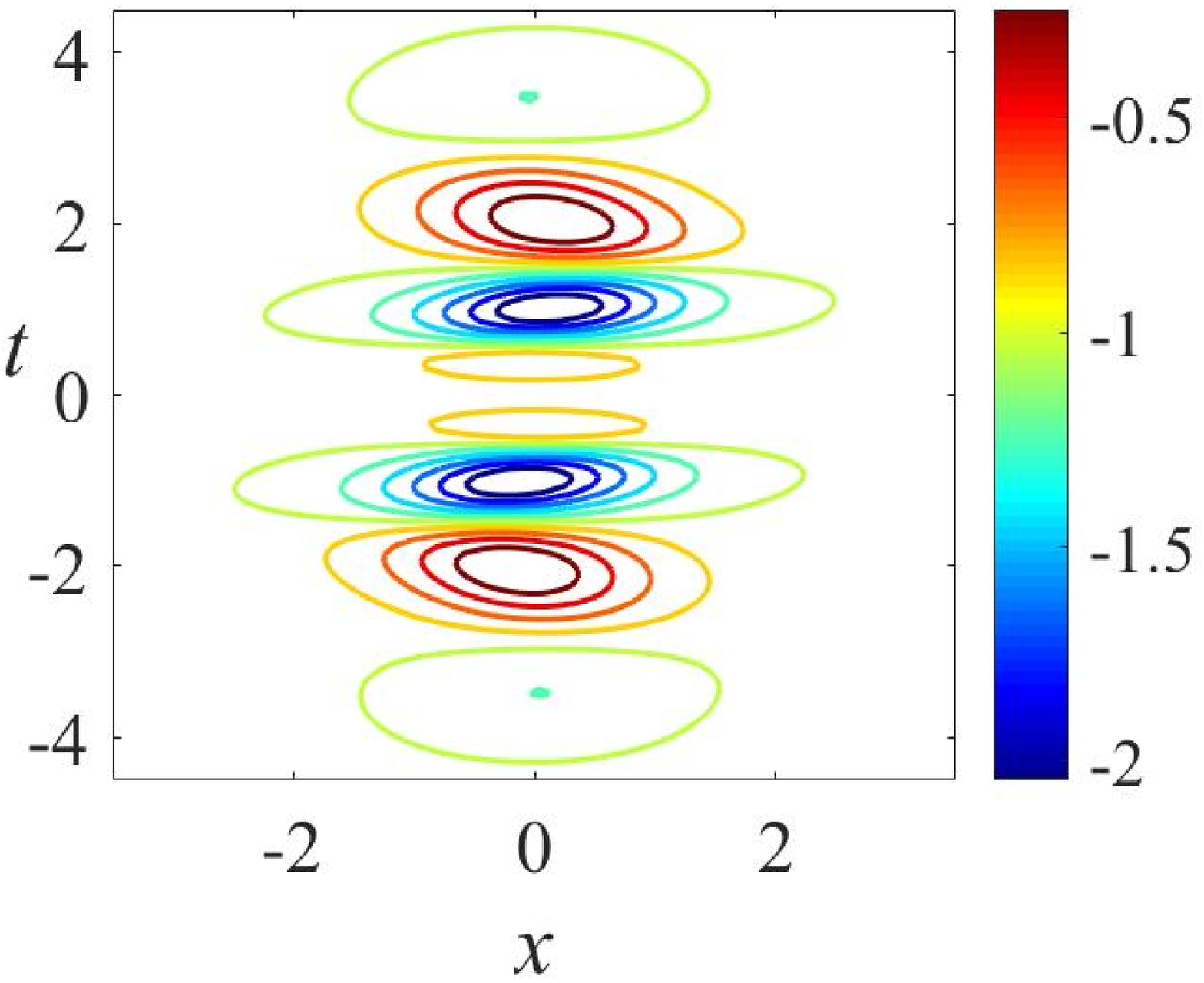}
\end{minipage}%
}
\subfigure[]{
\begin{minipage}[t]{0.43\linewidth}
\centering
\includegraphics[width=5.5cm]{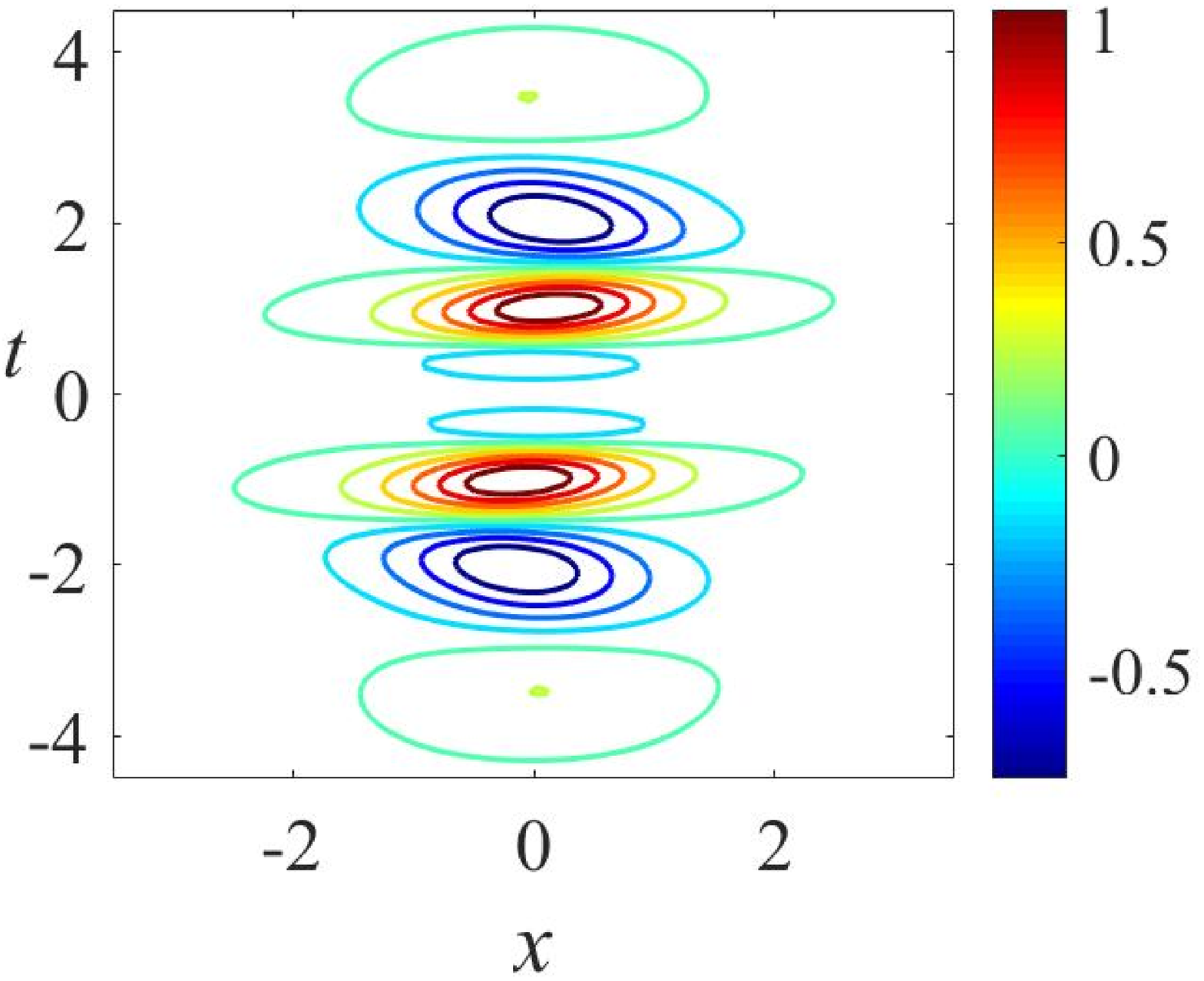}
\end{minipage}
}
\caption{Propagation of multipole instanton (15) with $k_{0}=1,\omega_{0}=2$: (a) spatial structure of $H^{y}$, (b) spatial structure of $H^{z}$, (c) contour plot of $H^{y}$, (d) contour plot of $H^{z}$.}
\end{figure}
\section{Rogue wave and interaction solutions}
\subsection{Rogue wave of the original KMM system}
In this section, we will study rogue wave solution and interaction solutions between rogue wave and multi-soliton for the original KMM system (2). Based on Painlev$\acute{\rm e}$ analysis, the Painlev$\acute{\rm e}$-B\"acklund transformation of the original KMM system (2) has the form
\begin{equation}
\begin{split}
&u=\frac{u_{0}}{\phi^{2}}+\frac{u_{1}}{\phi}+u_{2},\ \ \ \ v=\frac{v_{0}}{\phi}+v_{1},
\end{split}
\end{equation}
where $\phi,u_{0},u_{1},u_{2},v_{0},v_{1}$ are arbitrary functions of $x$ and $t$. For the sake of simplicity, we choose the seed solution $u_{2}=0$. Substituting (16) into (2) and setting the coefficients of $\phi^{-3}$ equal to zero, $u_{0}$ and $v_{0}$ are supposed to be expressed as
\begin{equation}
\begin{split}
&u_{0}=2\phi_{x}\phi_{t},\ \ \ \ v_{0}=2\phi_{t},
\end{split}
\end{equation}
vanishing the coefficients of $\phi^{-2}$, we obtain
\begin{equation}
\begin{split}
&u_{1}=-2\phi_{xt},\ \ \ \ v_{1}=-\frac{\phi_{tt}}{\phi_{t}},
\end{split}
\end{equation}
and the solutions of Eq.(2)
\begin{equation}
\begin{split}
&u=\frac{2\phi_{x}\phi_{t}}{\phi^{2}}-\frac{2\phi_{xt}}{\phi},\ \ \ \ v=\frac{2\phi_{t}}{\phi}-\frac{\phi_{tt}}{\phi_{t}}.
\end{split}
\end{equation}
By substituting (19) into (2), a quadri-linear form coupled with a tri-linear form for the original KMM system is obtained
\begin{equation}
\begin{split}
-\phi\phi_{t}^{2}\phi_{xttt}+\phi\phi_{t}\phi_{xt}\phi_{ttt}+2\phi\phi_{t}&\phi_{xtt}\phi_{tt}-2\phi\phi_{xt}\phi_{tt}^{2}+2\phi_{t}^{3}\phi_{xtt}-2\phi_{t}^{2}\phi_{xt}\phi_{tt}=0,\\
&\phi_{tt}\left(\phi_{t}\phi_{xtt}-\phi_{xt}\phi_{tt}\right)=0.
\end{split}
\end{equation}
Concretely, for a given solution of the coupled equations (20), the expression (19) will yield a corresponding exact solution to the KMM system (2) directly.

To search for rogue wave of the original KMM system, we assume the form of function $\phi$ as
\begin{equation}
\begin{split}
\phi=\sum_{i=1}^{2}\xi_{i}^{2}+a_{0},\ \ \ \ \xi_{i}=k_{i}x+\omega_{i}t+l_{i},
\end{split}
\end{equation}
where $k_{i},\omega_{i},l_{i}(1\le i\le 2)$ and $a_{0}$ are real parameters to be determined. Substituting expression (21) into (20) and vanishing the coefficients of all powers of $x$ and $t$, a simple constraining equation for the parameters is obtained
\begin{equation}
\begin{split}
k_{1}\omega_{1}+k_{2}\omega_{2}=0,
\end{split}
\end{equation}
which leads to the quadratic solutions to Eq.(20) as follows
\begin{equation}
\begin{split}
&\phi=(k_{1}x-\frac{k_{2}\omega_{2}}{k_{1}}t+l_{1})^{2}+(k_{2}x+\omega_{2}t+l_{2})^{2}+a_{0}.\\
\end{split}
\end{equation}
A class of rogue wave solutions for the original KMM system are obtained via (19)
\begin{equation}
\begin{split}
u=\frac{8(k_{1}\xi_{1}+k_{2}\xi_{2})(\omega_{1}\xi_{1}+\omega_{2}\xi_{2})}{(\xi_{1}^{2}+\xi_{2}^{2}+a_{0})^{2}},\ \ \ \ v=\frac{4(\omega_{1}\xi_{1}+\omega_{2}\xi_{2})}{\xi_{1}^{2}+\xi_{2}^{2}+a_{0}}-\frac{\omega_{1}^{2}+\omega_{2}^{2}}{\omega_{1}\xi_{1}+\omega_{2}\xi_{2}},
\end{split}
\end{equation}
which needs to satisfy $k_{1}\ne0,a_{0}>0$ (this conditions are required in all of solutions obtained in this section) to guarantee the well-definedness of $\phi$. Correspondingly, it is not hard to present the two components of magnetic field $H^{y}$ and $H^{z}$ from $u$ and $v$. Fig.4 displays a special rogue wave with the parameters selected as $k_{1}=-2/3,l_{1}=1,k_{2}=2,\omega_{2}=1/3,l_{2}=2,a_{0}=1$, respectively. The spatial structures of $H^{y}$ and $H^{z}$ are described in Figs.4(a) and 4(d). Figs.4(b) and 4(e) represent the density plots of the rogue wave. Figs.4(c) and 4(f) show the contour plots of the rogue wave. It clearly appears that the rogue wave, which only occurs in $-10\le x\le 10$,$-10\le t\le 10$, originates from the background energy of $H^{y}$ and transmits energy to $H^{z}$ through the nonlinear interaction.
\begin{figure}[htbp]
\subfigure[]{
\begin{minipage}[t]{0.33\linewidth}
\centering
\includegraphics[width=4.5cm]{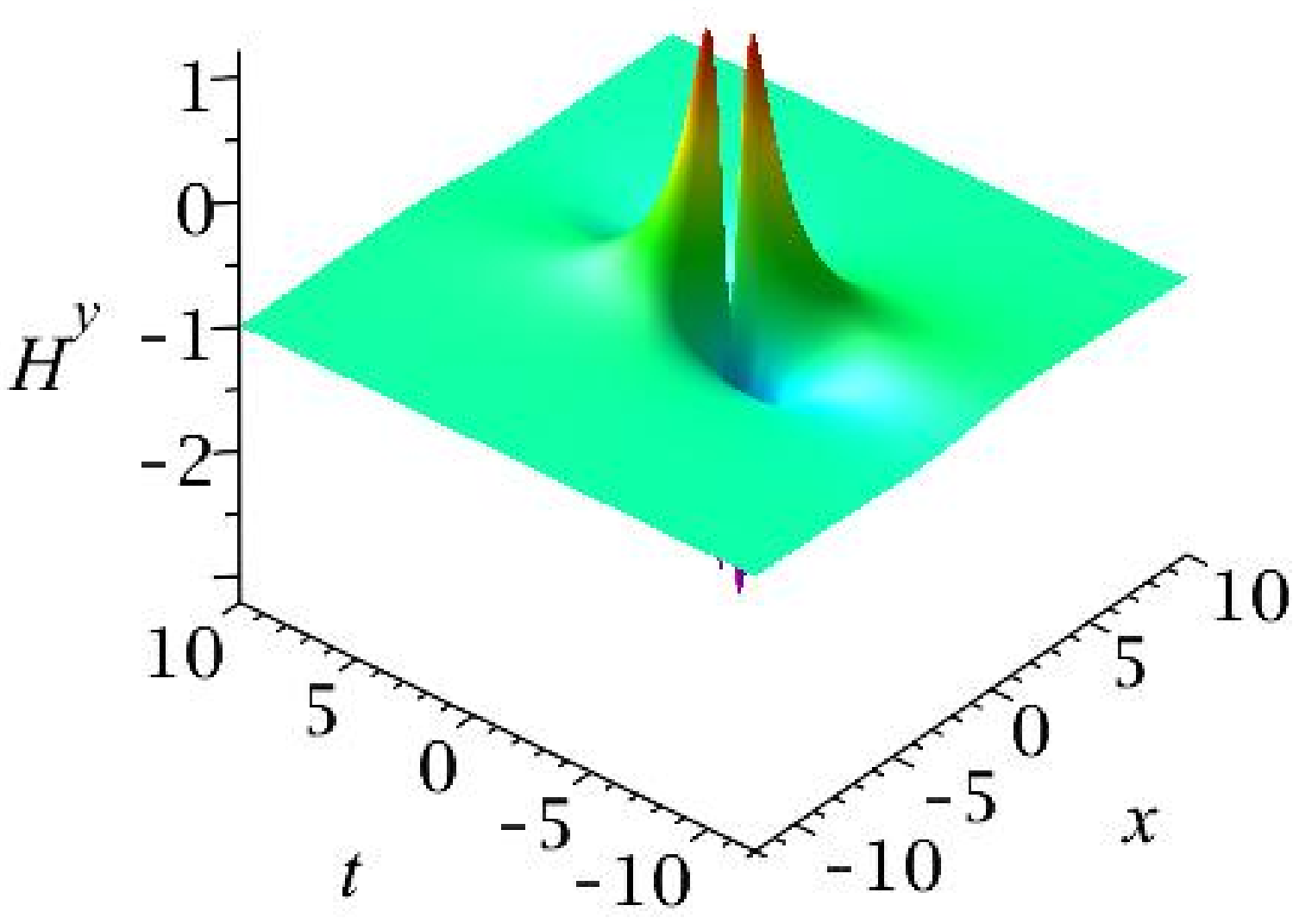}
\end{minipage}%
}
\subfigure[]{
\begin{minipage}[t]{0.33\linewidth}
\centering
\includegraphics[width=4.75cm]{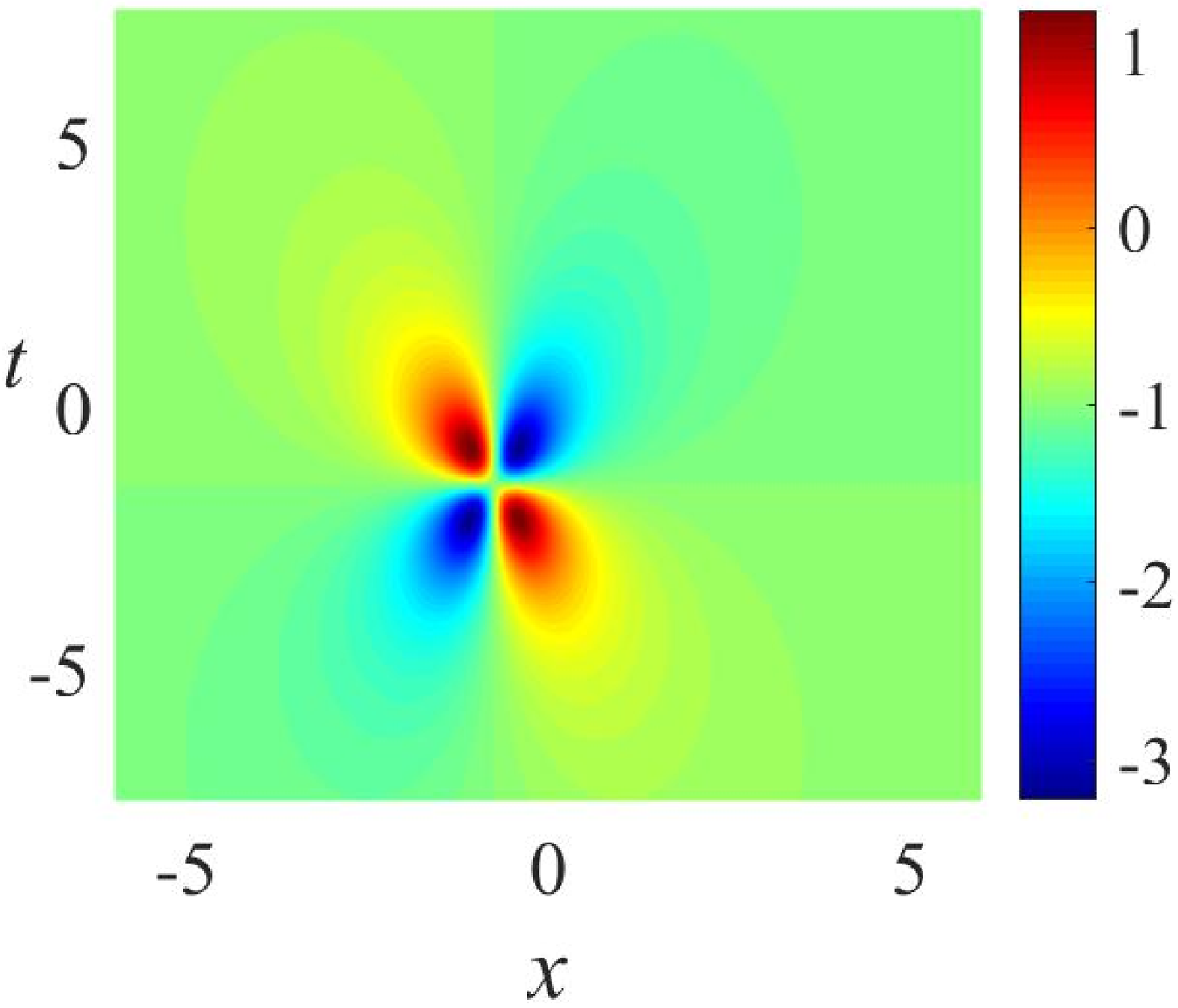}
\end{minipage}
}
\subfigure[]{
\begin{minipage}[t]{0.33\linewidth}
\centering
\includegraphics[width=4.75cm]{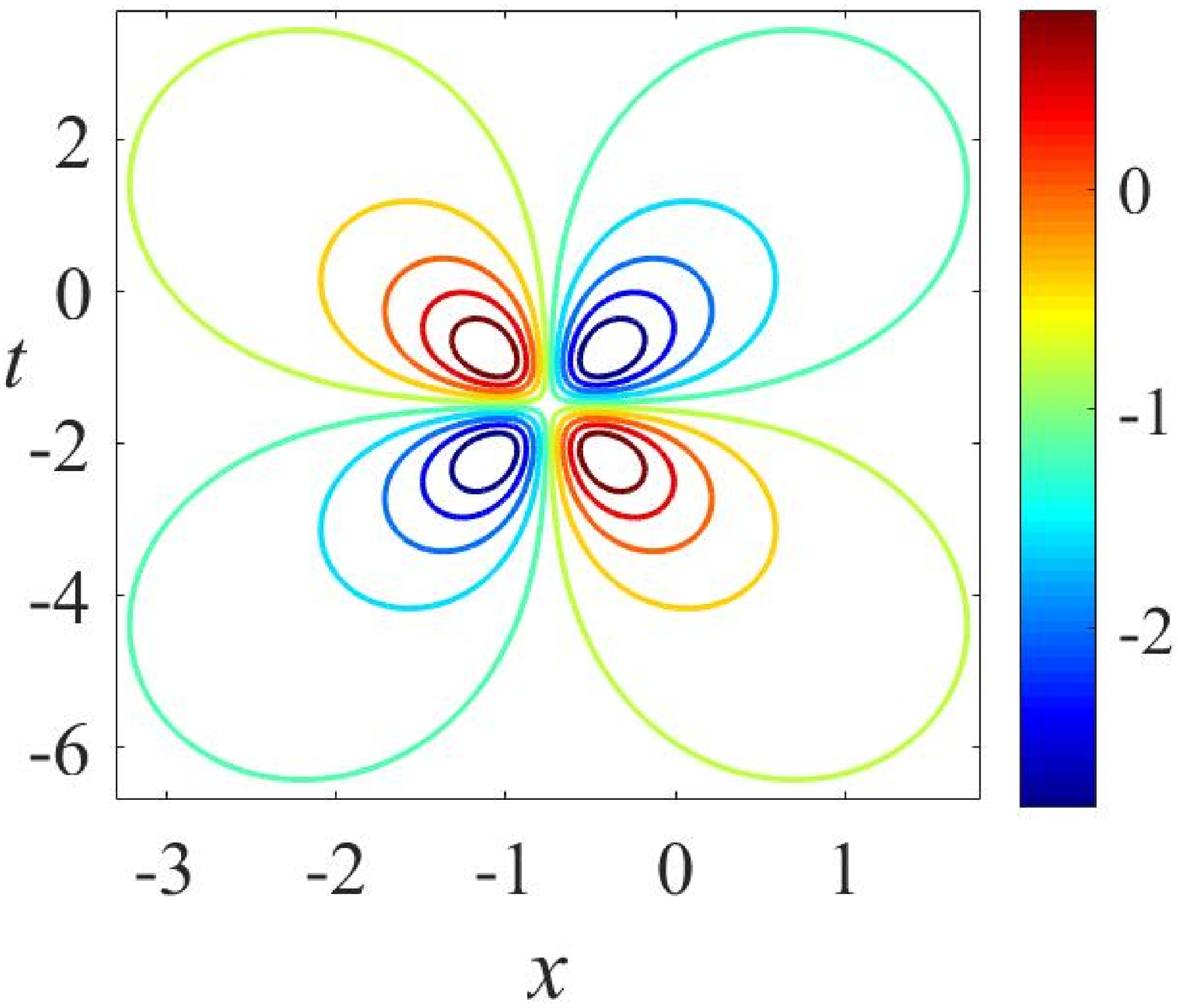}
\end{minipage}%
}

\subfigure[]{
\begin{minipage}[t]{0.33\linewidth}
\centering
\includegraphics[width=4.5cm]{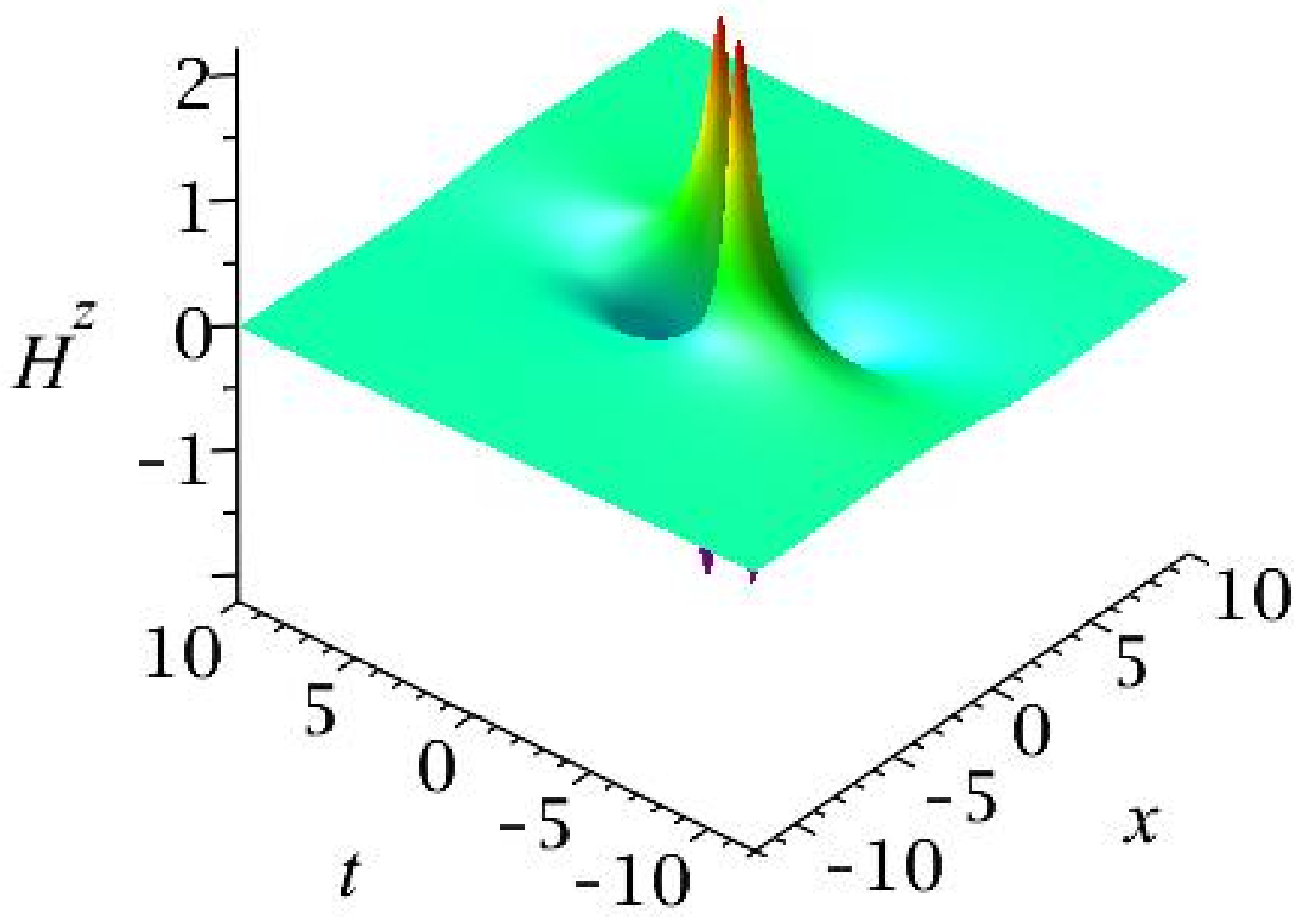}
\end{minipage}
}
\subfigure[]{
\begin{minipage}[t]{0.33\linewidth}
\centering
\includegraphics[width=4.75cm]{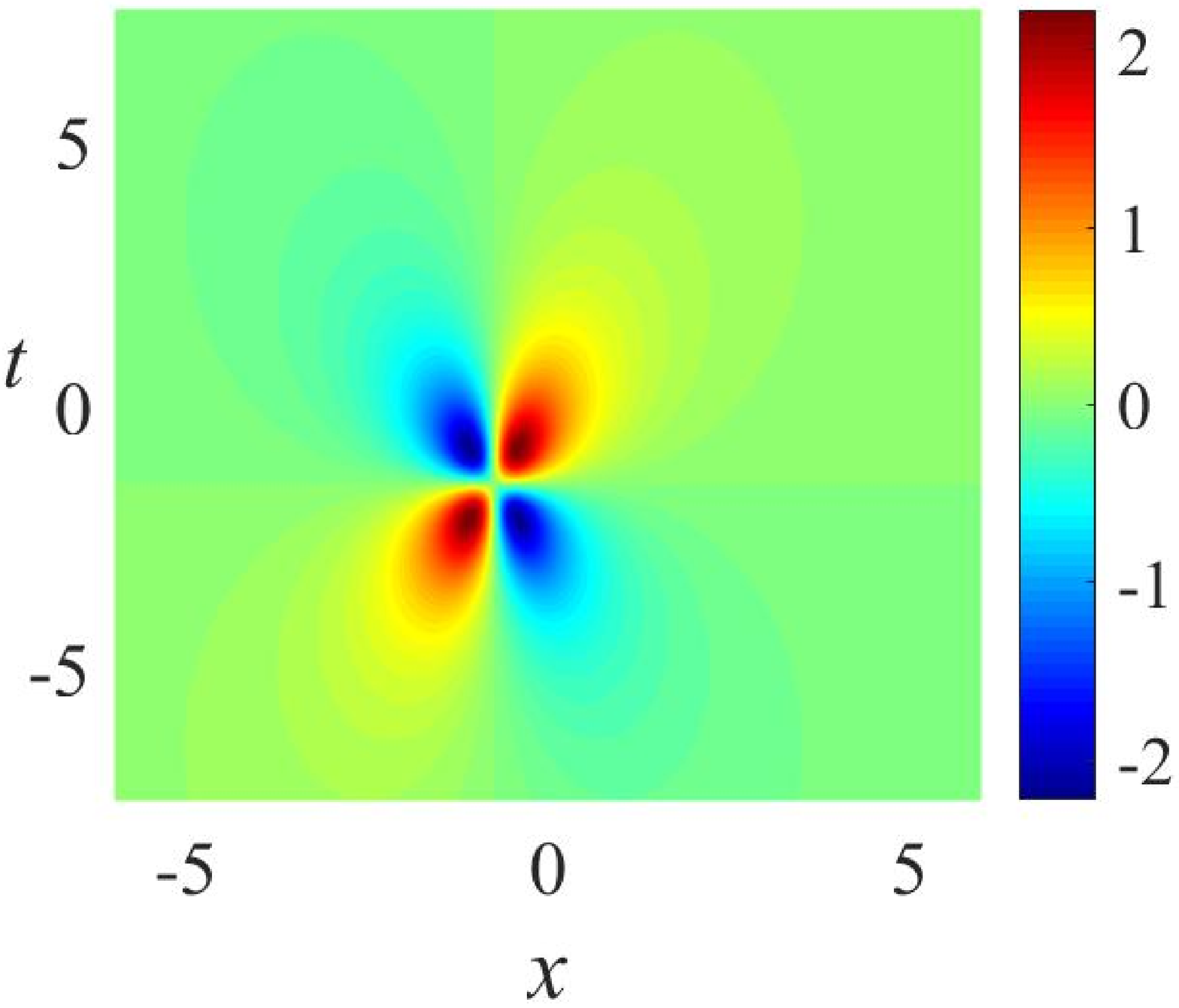}
\end{minipage}%
}
\subfigure[]{
\begin{minipage}[t]{0.33\linewidth}
\centering
\includegraphics[width=4.75cm]{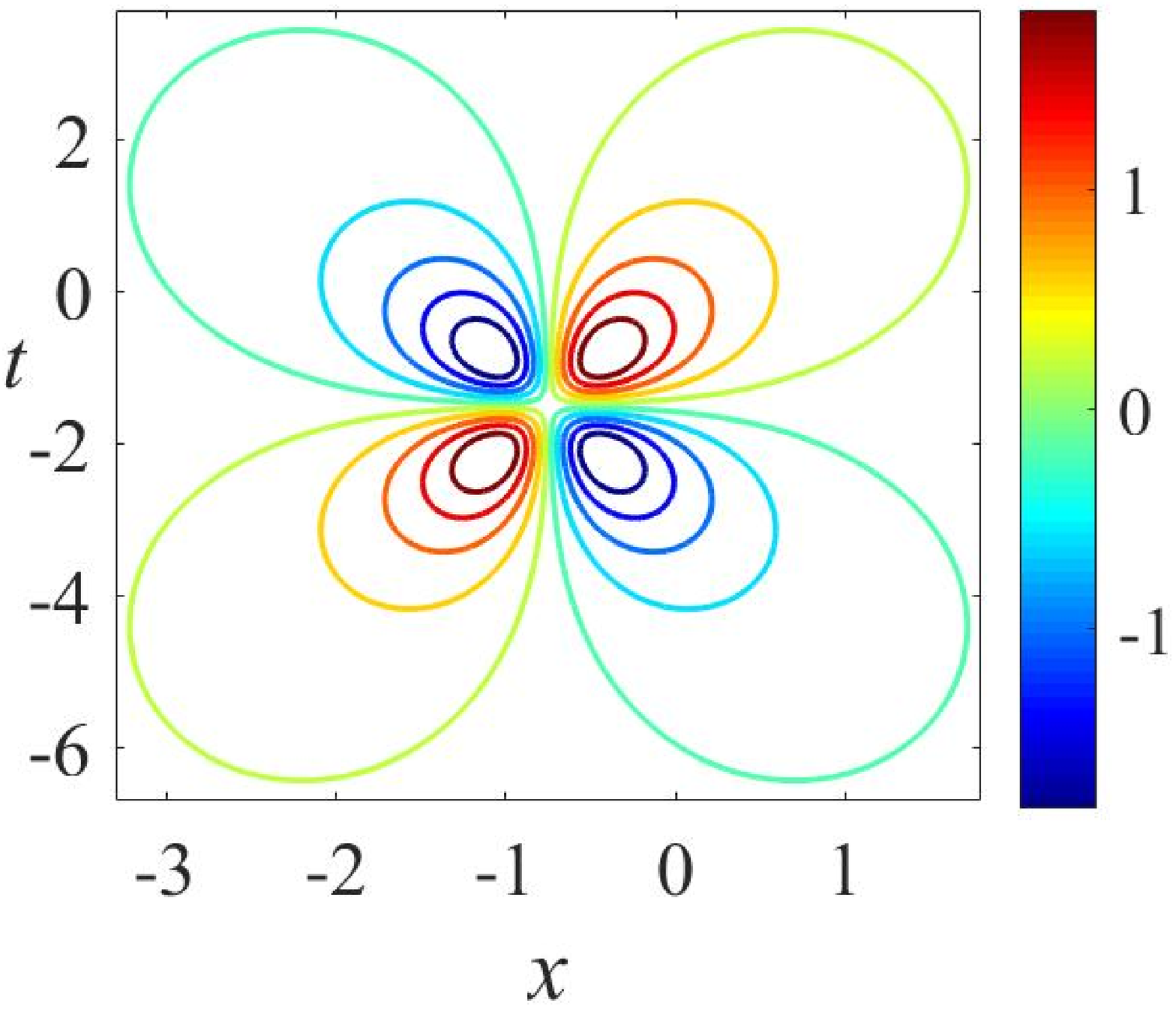}
\end{minipage}
}
\caption{Propagation of rogue wave (24): (a) the evolution plot of $H^{y}$, (b) the density plot of $H^{y}$, (c) the contour plot of $H^{y}$, (d) the evolution plot of $H^{z}$, (e) the density plot of $H^{z}$, (f) the contour plot of $H^{z}$.}
\end{figure}
\subsection{Interaction solutions between Rogue wave and one-soliton}
Interaction solutions between rogue wave and other type solutions can be obtained by combining the quadratic function with additional functions \cite{39,40}. In order to find interaction solutions between rogue wave and one-soliton, we assume solutions of (20) as a sum of a quadratic function and an exponential function \cite{41}
\begin{equation}
\begin{split}
\phi=\sum_{i=1}^{2}\xi_{i}^{2}+a_{0}+e^{\xi_{3}},\ \ \ \ \xi_{i}=k_{i}x+\omega_{i}t+l_{i},
\end{split}
\end{equation}
with $k_{i},\omega_{i},l_{i}(1\le i\le 3)$ and $a_{0}$ being undetermined real parameters. By substituting expression (25) into (20) and vanishing the coefficients of all powers of $x$ and $t$, two sets of constraining equations for parameters are obtained.

\textbf{\emph{Case I.}}\ \ $\omega_{1}=-\frac{k_{2}\omega_{2}}{k_{1}}$ and $k_{3}=0$

The interaction solutions between rogue wave and single soliton is
\begin{equation}
\begin{split}
u=\frac{4(k_{1}\xi_{1}+k_{2}\xi_{2})(2\omega_{1}\xi_{1}+2\omega_{2}\xi_{2}+\omega_{3}e^{\xi_{3}})}{(\xi_{1}^{2}+\xi_{2}^{2}+a_{0}+e^{\xi_{3}})^{2}},\ \ \ \ v=\frac{2(2\omega_{1}\xi_{1}+2\omega_{2}\xi_{2}+\omega_{3}e^{\xi_{3}})}{\xi_{1}^{2}+\xi_{2}^{2}+a_{0}+e^{\xi_{3}}}-\frac{2\omega_{1}^{2}+2\omega_{2}^{2}+\omega_{3}^{2}e^{\xi_{3}}}{2\omega_{1}\xi_{1}+2\omega_{2}\xi_{2}+\omega_{3}e^{\xi_{3}}}.
\end{split}
\end{equation}
The propagation plot of the interaction solutions between rogue wave and single soliton is like a butterfly-type structure (See Fig.5).
\begin{figure}[htbp]
\subfigure[]{
\begin{minipage}[t]{0.33\linewidth}
\centering
\includegraphics[width=4.5cm]{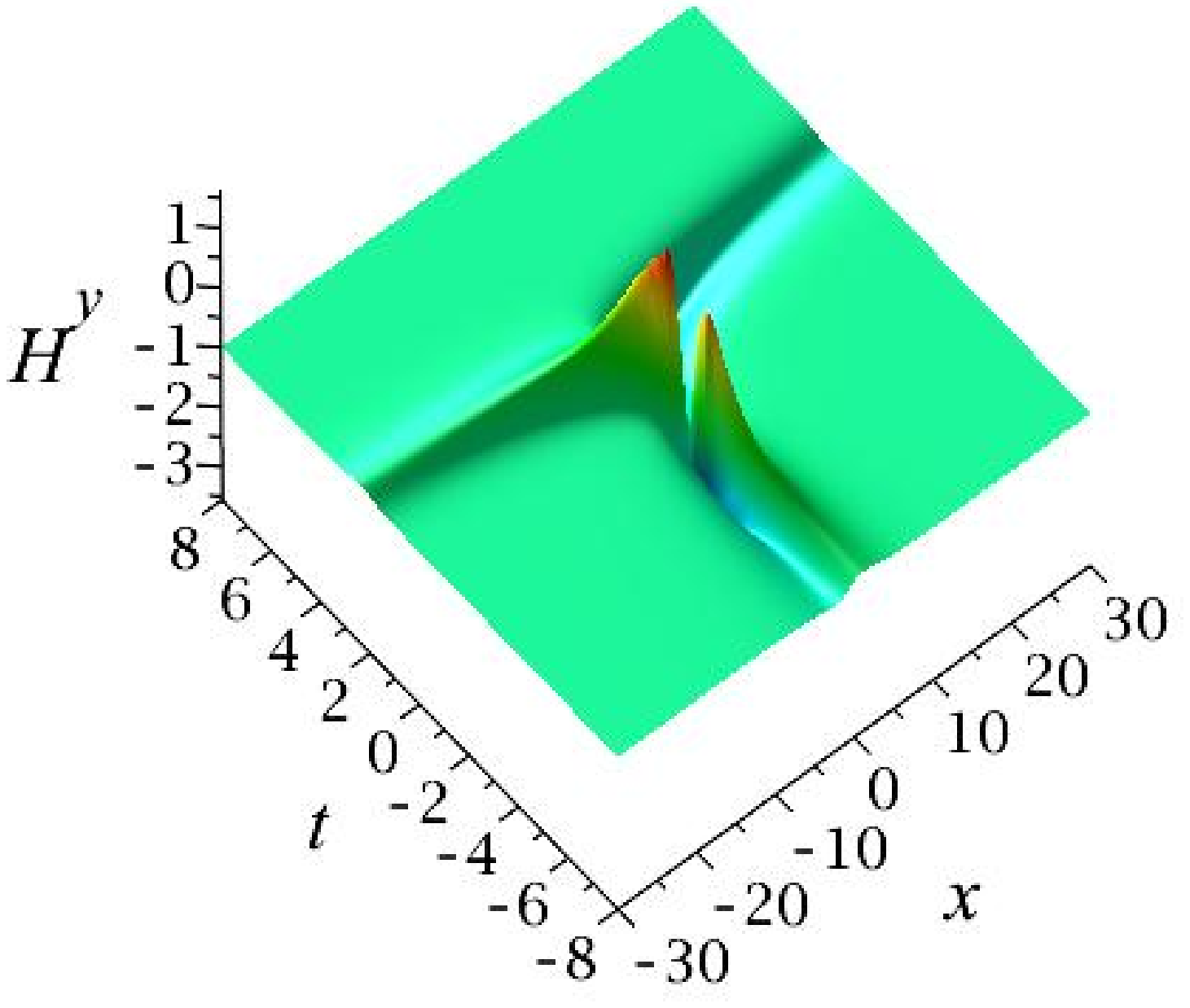}
\end{minipage}%
}
\subfigure[]{
\begin{minipage}[t]{0.33\linewidth}
\centering
\includegraphics[width=4.75cm]{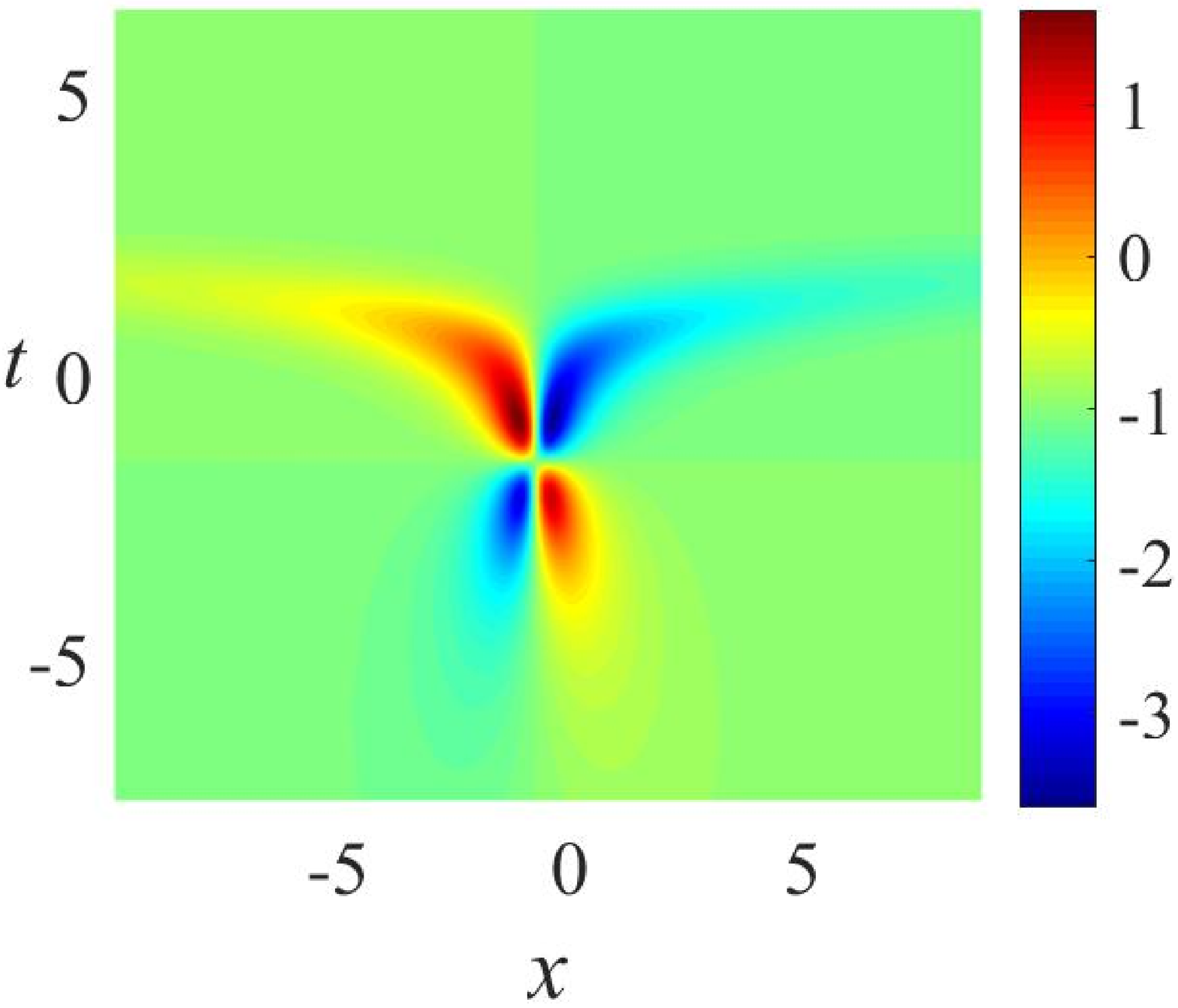}
\end{minipage}
}
\subfigure[]{
\begin{minipage}[t]{0.33\linewidth}
\centering
\includegraphics[width=4.75cm]{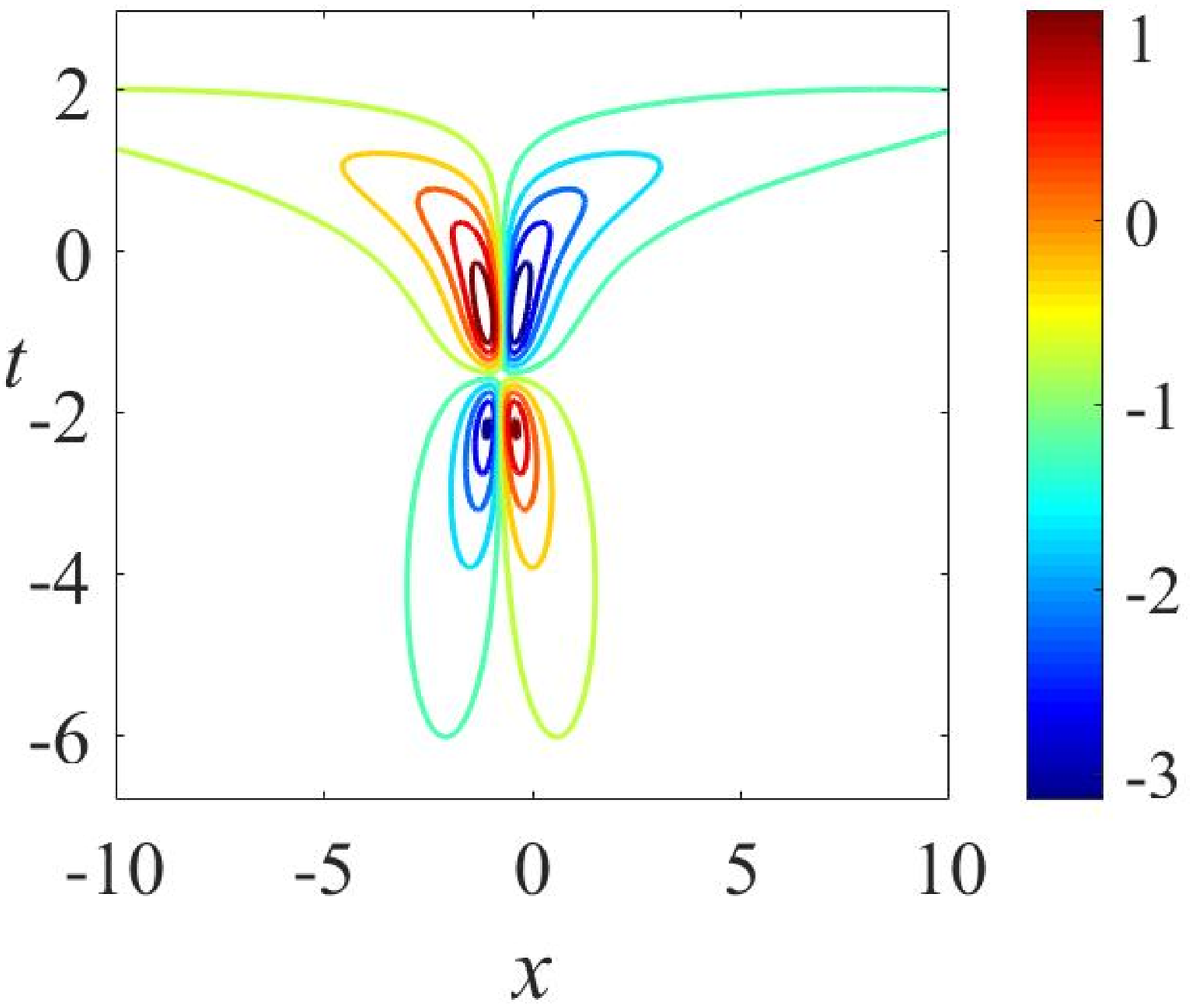}
\end{minipage}%
}

\subfigure[]{
\begin{minipage}[t]{0.33\linewidth}
\centering
\includegraphics[width=4.5cm]{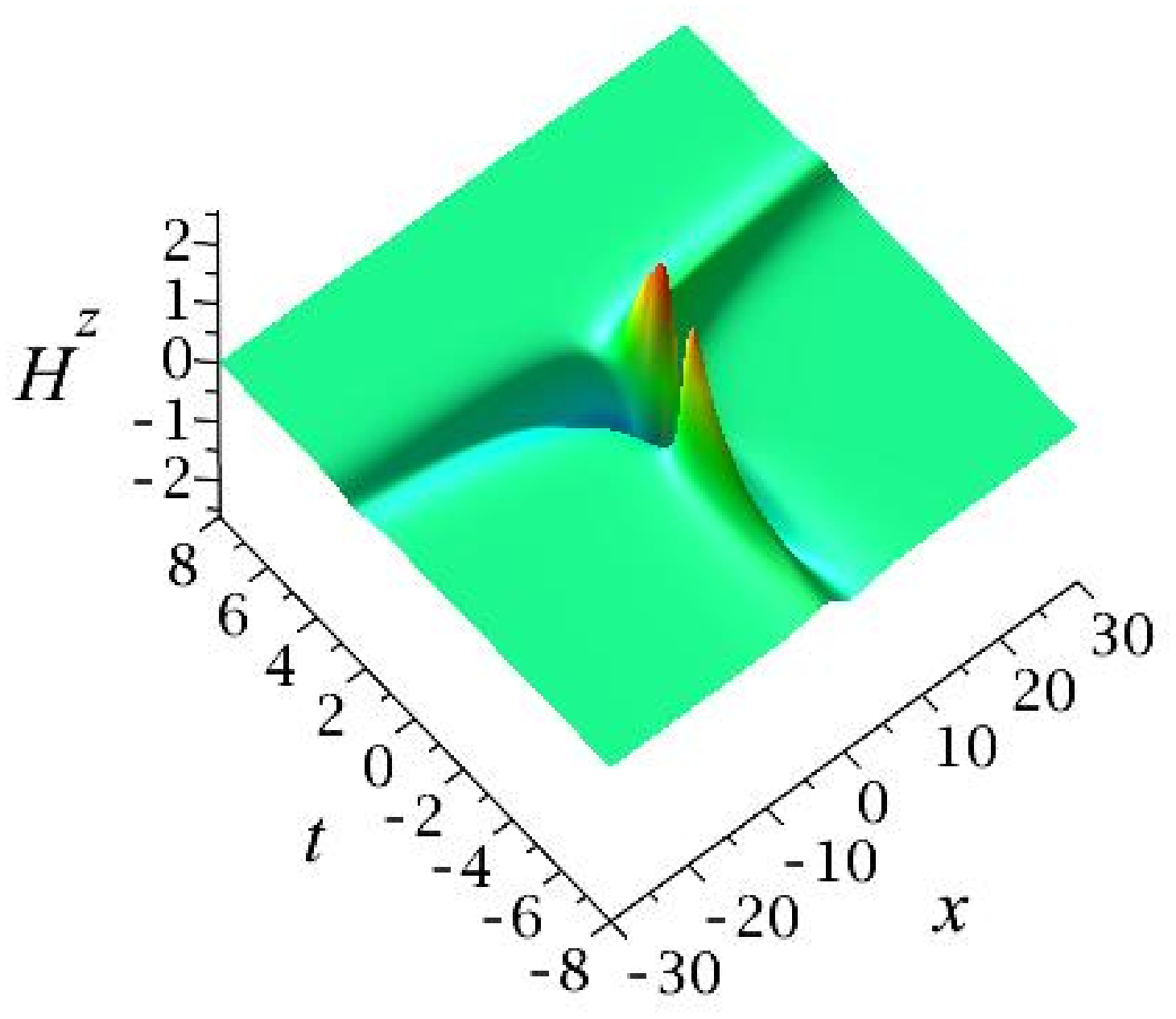}
\end{minipage}
}
\subfigure[]{
\begin{minipage}[t]{0.33\linewidth}
\centering
\includegraphics[width=4.75cm]{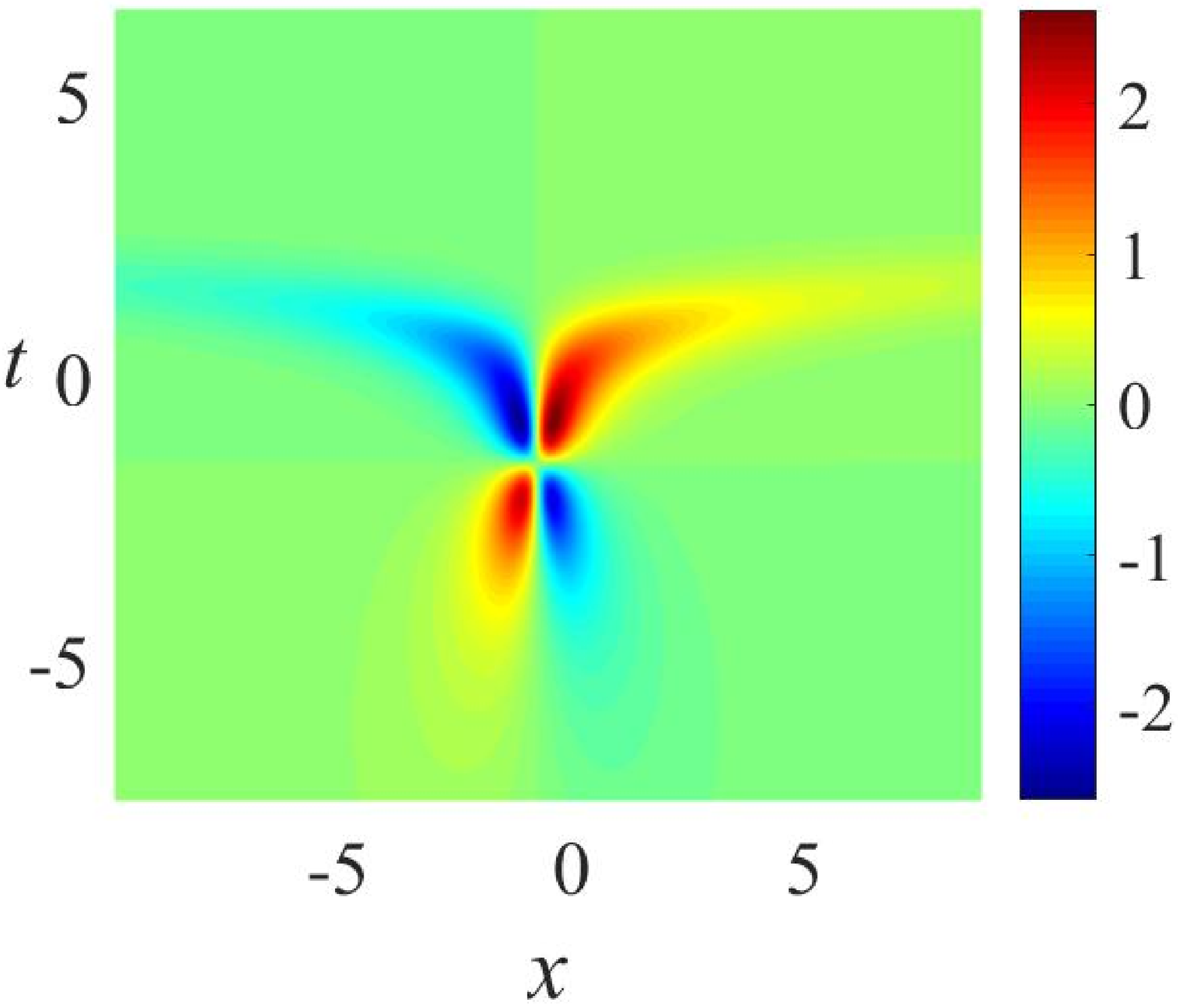}
\end{minipage}%
}
\subfigure[]{
\begin{minipage}[t]{0.33\linewidth}
\centering
\includegraphics[width=4.75cm]{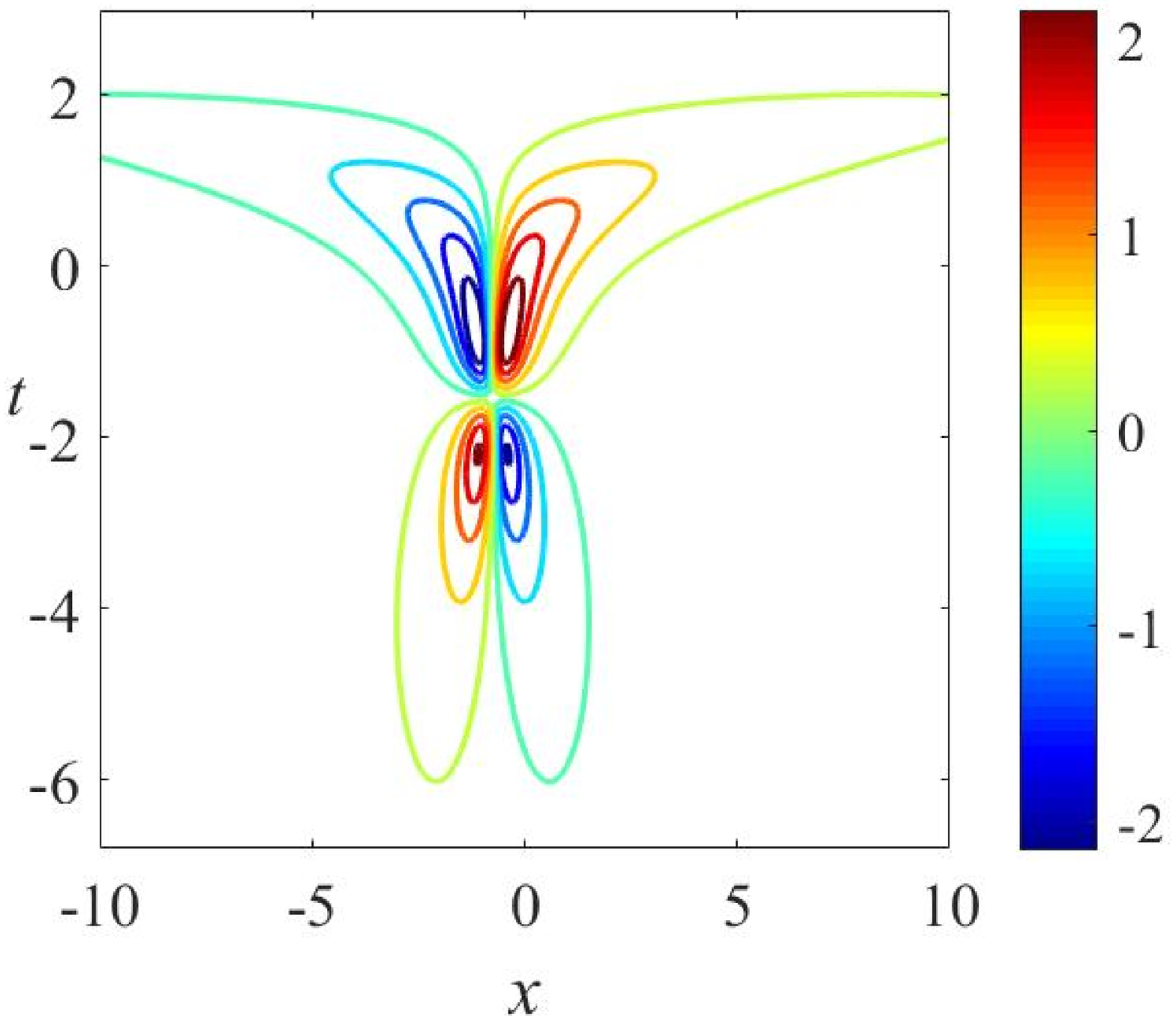}
\end{minipage}
}
\caption{Propagation of butterfly-shaped solution (26) of $H^{y}$ and $H^{z}$ with the parameters selected as $k_{1}=-2/3,l_{1}=1,k_{2}=2,\omega_{2}=1/3,l_{2}=2,a_{0}=1,\omega_{3}=3,l_{3}=1$: (a) evolution plot of $H^{y}$, (b) density plot of $H^{y}$, (c) contour plot of $H^{y}$, (d) evolution plot of $H^{z}$, (e) density plot of $H^{z}$, (f) contour plot of $H^{z}$.}
\end{figure}

\textbf{\emph{Case II.}}\ \ $\omega_{1}=-\frac{k_{2}\omega_{2}}{k_{1}}$ and $\omega_{3}=0$

The second type interaction solution between rogue wave and single soliton read
\begin{equation}
\begin{split}
u=\frac{4(2k_{1}\xi_{1}+2k_{2}\xi_{2}+k_{3}e^{\xi_{3}})(\omega_{1}\xi_{1}+\omega_{2}\xi_{2})}{(\xi_{1}^{2}+\xi_{2}^{2}+a_{0}+e^{\xi_{3}})^{2}},\ \ \ \ v=\frac{4(\omega_{1}\xi_{1}+\omega_{2}\xi_{2})}{\xi_{1}^{2}+\xi_{2}^{2}+a_{0}+e^{\xi_{3}}}-\frac{\omega_{1}^{2}+\omega_{2}^{2}}{\omega_{1}\xi_{1}+\omega_{2}\xi_{2}}
\end{split}
\end{equation}
This interactional property is exactly the same as butterfly-type solution (26) except for transmission direction.
\subsection{Interaction solutions between Rogue wave and mixed solitons}
We also can derive the interaction solution between rogue wave and mixed solitons. By taking the form of a quadratic function and two exponential functions:
\begin{equation}
\begin{split}
\phi=\sum_{i=1}^{2}\xi_{i}^{2}+a_{0}+\sum_{i=3}^{4}e^{\xi_{i}},\ \ \ \ \xi_{i}=k_{i}x+\omega_{i}t+l_{i},
\end{split}
\end{equation}
with $k_{i},\omega_{i},l_{i}(1\le i\le 4) \ \rm{and} \ a_{0}$ being undetermined real parameters. Substituting Eq.(31) into Eq.(20) and vanishing coefficients of $x$, $t$ and exponential function, we find two cases of solutions, without losing generality.

\textbf{\emph{Case I.}}\ \ $\omega_{1}=-\frac{k_{2}\omega_{2}}{k_{1}}$ and $k_{3}=k_{4}=0$

The X-type interaction solution of Eq.(2) reads
\begin{equation}
\begin{split}
u=&\frac{4(k_{1}\xi_{1}+k_{2}\xi_{2})(2\omega_{1}\xi_{1}+2\omega_{2}\xi_{2}+\omega_{3}e^{\xi_{3}}+\omega_{4}e^{\xi_{4}})}{(\xi_{1}^{2}+\xi_{2}^{2}+a_{0}+e^{\xi_{3}}+e^{\xi_{4}})^{2}},\\ v=&\frac{2(2\omega_{1}\xi_{1}+2\omega_{2}\xi_{2}+\omega_{3}e^{\xi_{3}}+\omega_{4}e^{\xi_{4}})}{\xi_{1}^{2}+\xi_{2}^{2}+a_{0}+e^{\xi_{3}}+e^{\xi_{4}}}-\frac{2\omega_{1}^{2}+2\omega_{2}^{2}+\omega_{3}^{2}e^{\xi_{3}}+\omega_{4}^{2}e^{\xi_{4}}}{2\omega_{1}\xi_{1}+2\omega_{2}\xi_{2}+\omega_{3}e^{\xi_{3}}+\omega_{4}e^{\xi_{4}}},
\end{split}
\end{equation}
The parameters are selected as $k_{1}=-2/3,l_{1}=1,k_{2}=2,\omega_{2}=1/3,l_{2}=2,a_{0}=1,\omega_{3}=7,l_{3}=1,\omega_{4}=-7,l_{4}=-1$ in Fig.(6).

\textbf{\emph{Case II.}}\ \ $\omega_{1}=-\frac{k_{2}\omega_{2}}{k_{1}}$ and $k_{3}=\omega_{4}=0$

Another interaction solution of Eq.(2) is
\begin{equation}
\begin{split}
u=\frac{2(2k_{1}\xi_{1}+2k_{2}\xi_{2}+k_{4}e^{\xi_{4}})(2\omega_{1}\xi_{1}+2\omega_{2}\xi_{2}+\omega_{3}e^{\xi_{3}})}{(\xi_{1}^{2}+\xi_{2}^{2}+a_{0}+e^{\xi_{3}}+e^{\xi_{4}})^{2}},\ v=\frac{2(2\omega_{1}\xi_{1}+2\omega_{2}\xi_{2}+\omega_{3}e^{\xi_{4}})}{\xi_{1}^{2}+\xi_{2}^{2}+a_{0}+e^{\xi_{3}}+e^{\xi_{4}}}-\frac{2\omega_{1}^{2}+2\omega_{2}^{2}+\omega_{3}^{2}e^{\xi_{3}}}{2\omega_{1}\xi_{1}+2\omega_{2}\xi_{2}+\omega_{3}e^{\xi_{3}}}
\end{split}
\end{equation}
\begin{figure}[h]
\subfigure[]{
\begin{minipage}[t]{0.33\linewidth}
\centering
\includegraphics[width=4.6cm]{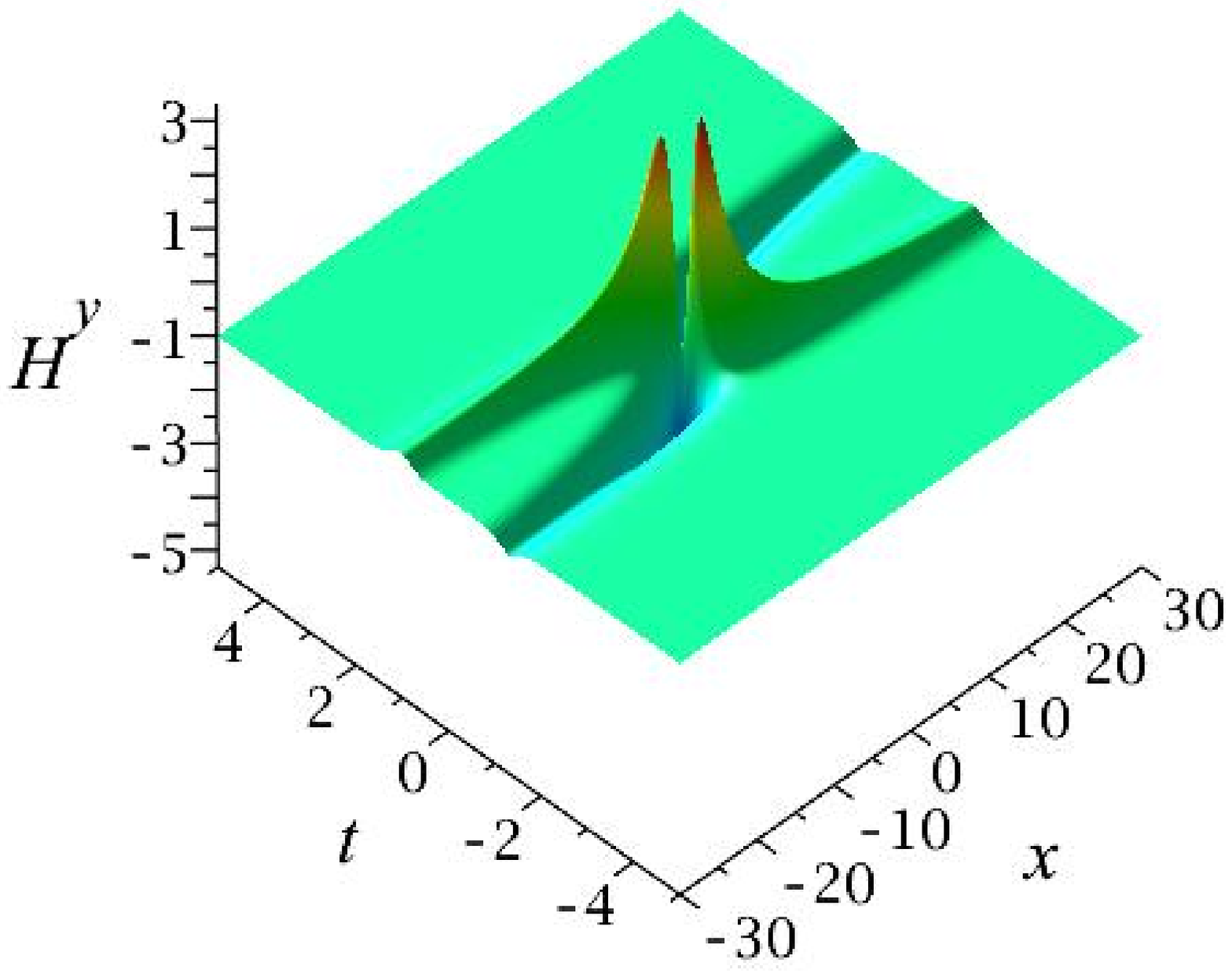}
\end{minipage}%
}
\subfigure[]{
\begin{minipage}[t]{0.33\linewidth}
\centering
\includegraphics[width=4.75cm]{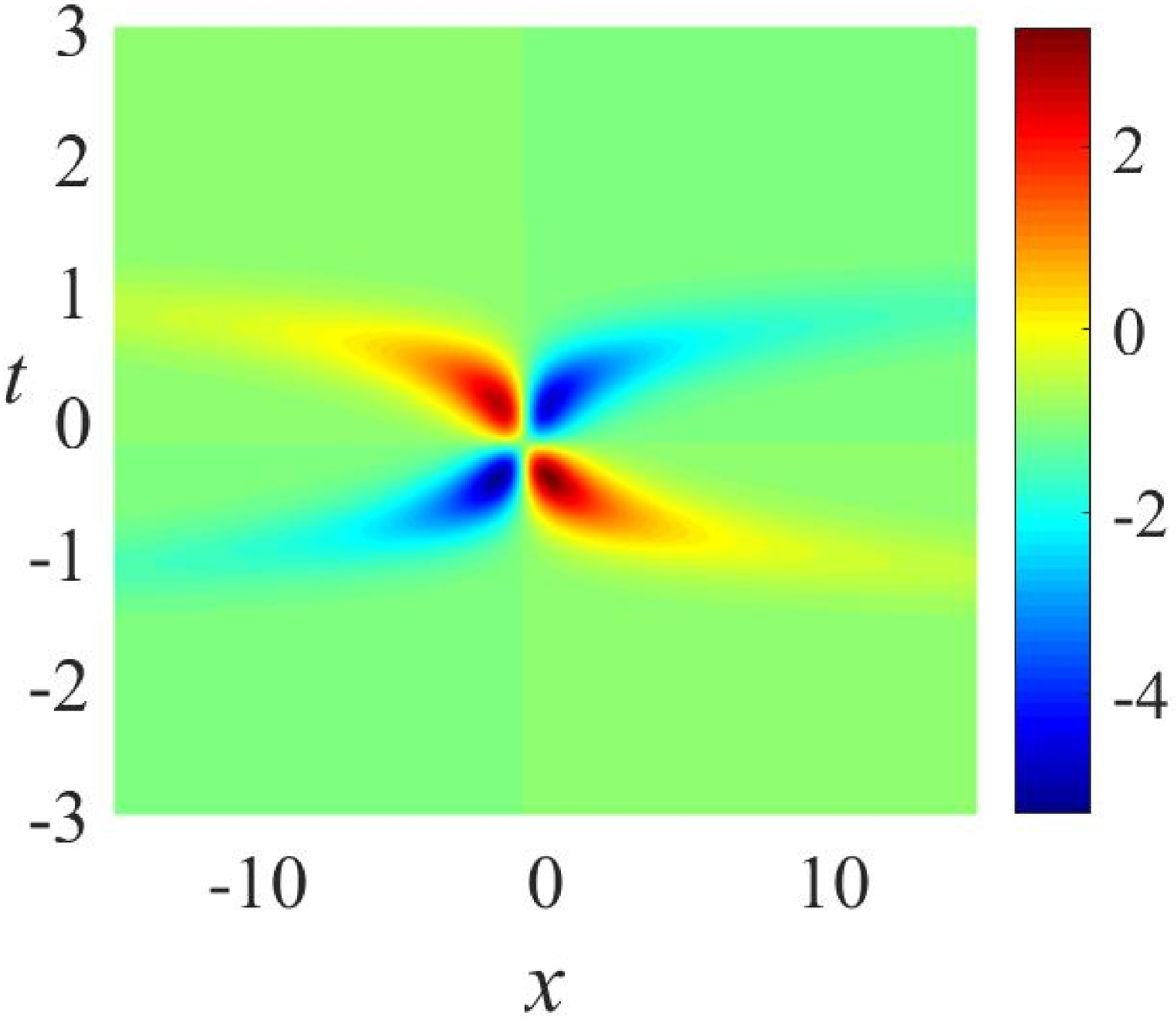}
\end{minipage}
}
\subfigure[]{
\begin{minipage}[t]{0.33\linewidth}
\centering
\includegraphics[width=4.75cm]{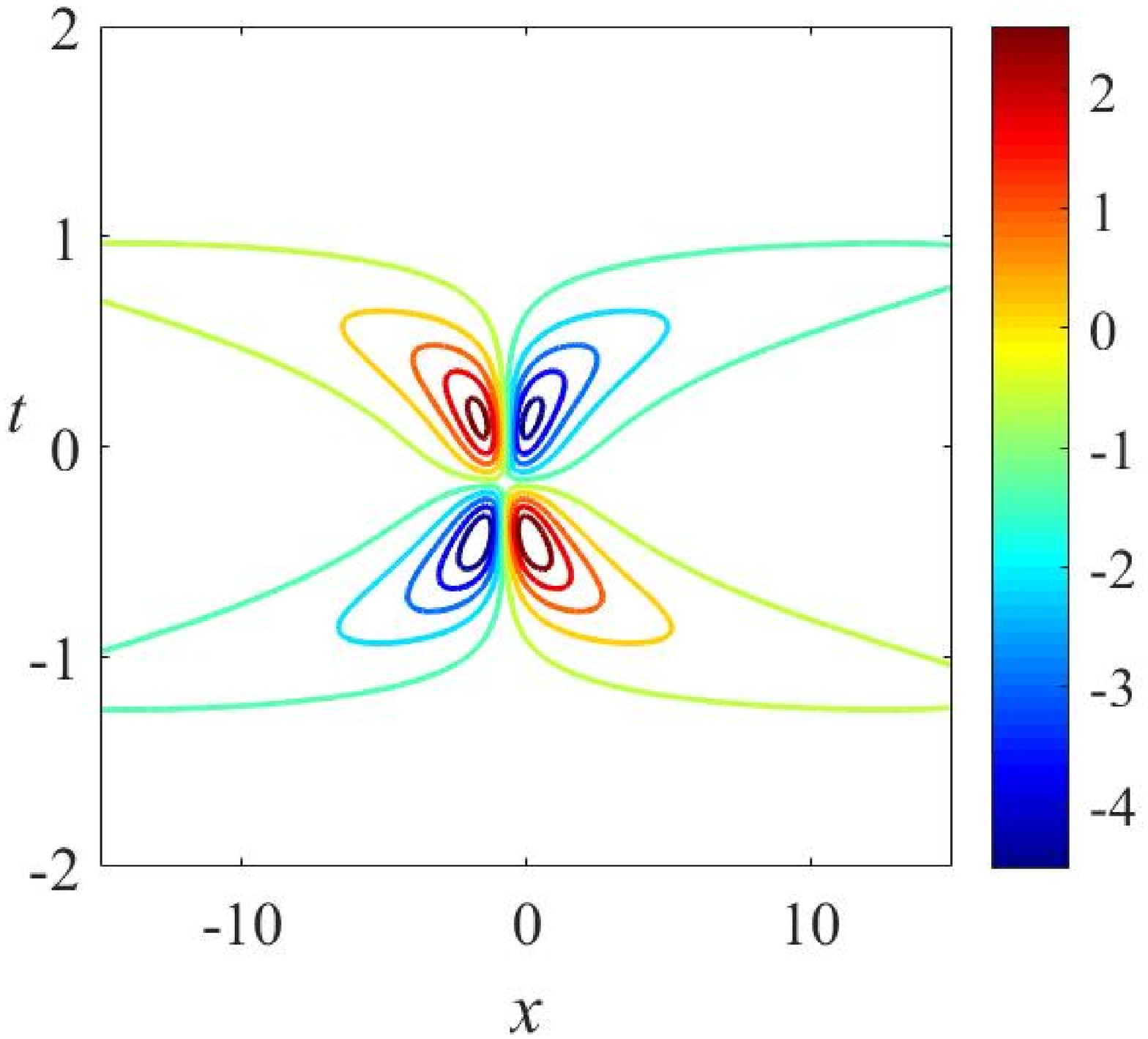}
\end{minipage}%
}

\subfigure[]{
\begin{minipage}[t]{0.33\linewidth}
\centering
\includegraphics[width=4.6cm]{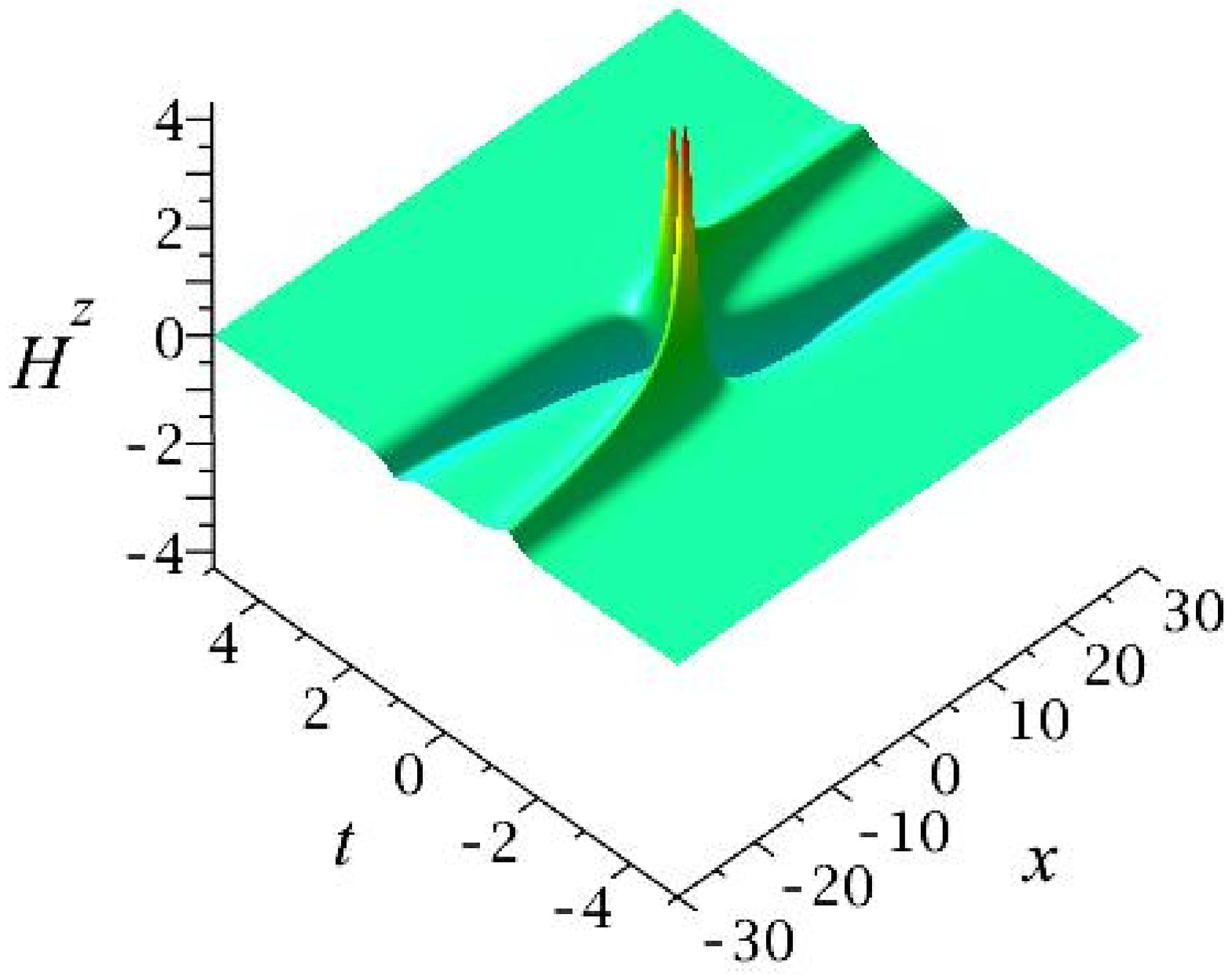}
\end{minipage}
}
\subfigure[]{
\begin{minipage}[t]{0.33\linewidth}
\centering
\includegraphics[width=4.75cm]{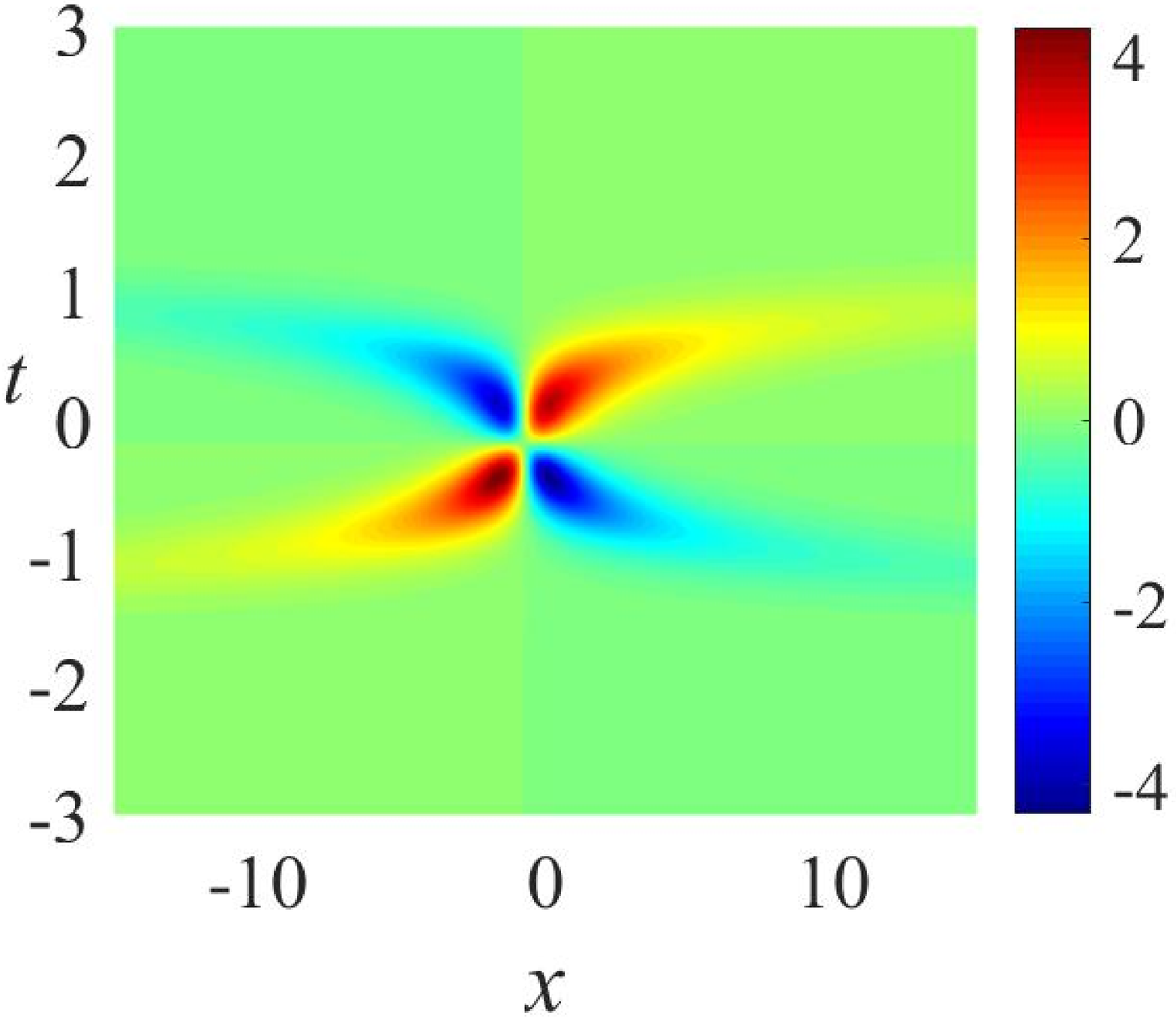}
\end{minipage}%
}
\subfigure[]{
\begin{minipage}[t]{0.33\linewidth}
\centering
\includegraphics[width=4.75cm]{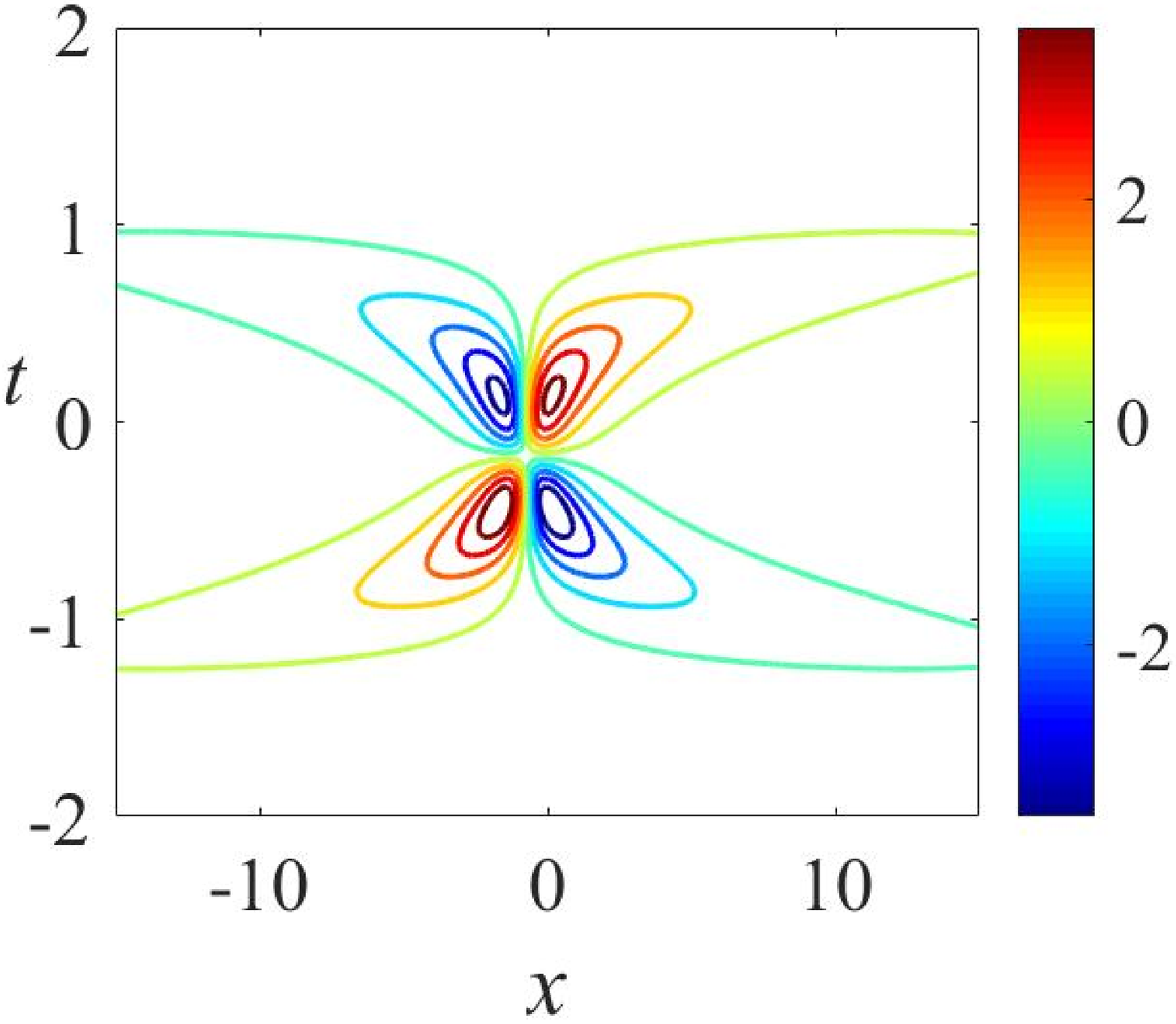}
\end{minipage}
}
\caption{The evolution plots to the X-type interaction solution (29) of $H^{y}$ and $H^{z}$ with $k_{1}=-2/3,l_{1}=1,k_{2}=2,\omega_{2}=1/3,l_{2}=2,a_{0}=1,\omega_{3}=7,l_{3}=1,\omega_{4}=-7,l_{4}=-1$: (a) evolution plot of $H^{y}$, (b) density plot of $H^{y}$, (c) contour plot of $H^{y}$, (d) evolution plot of $H^{z}$, (e) density plot of $H^{z}$, (f) contour plot of $H^{z}$.}
\end{figure}
\begin{figure}[h]
\subfigure[]{
\begin{minipage}[t]{0.33\linewidth}
\centering
\includegraphics[width=4.5cm]{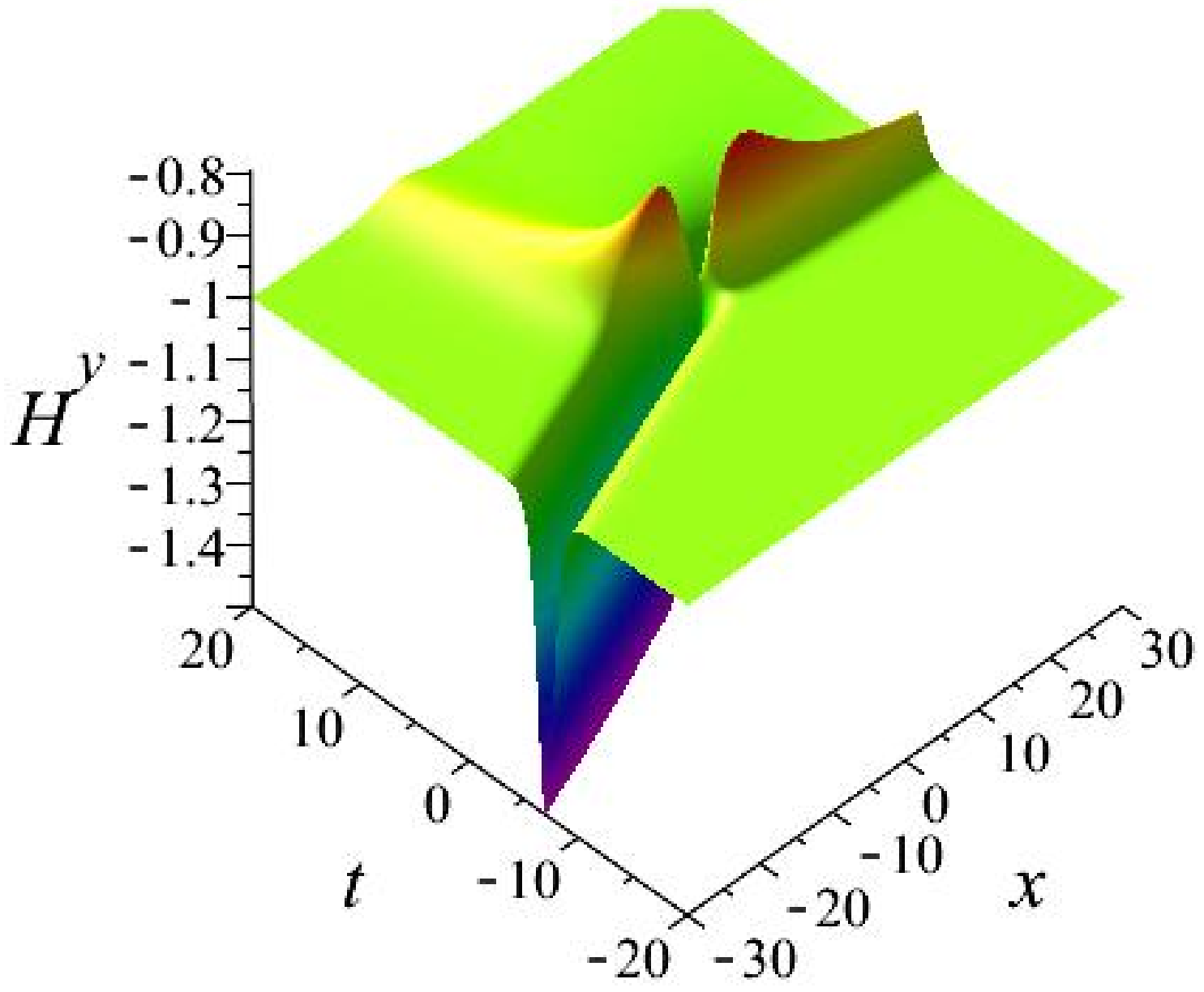}
\end{minipage}%
}
\subfigure[]{
\begin{minipage}[t]{0.33\linewidth}
\centering
\includegraphics[width=4.75cm]{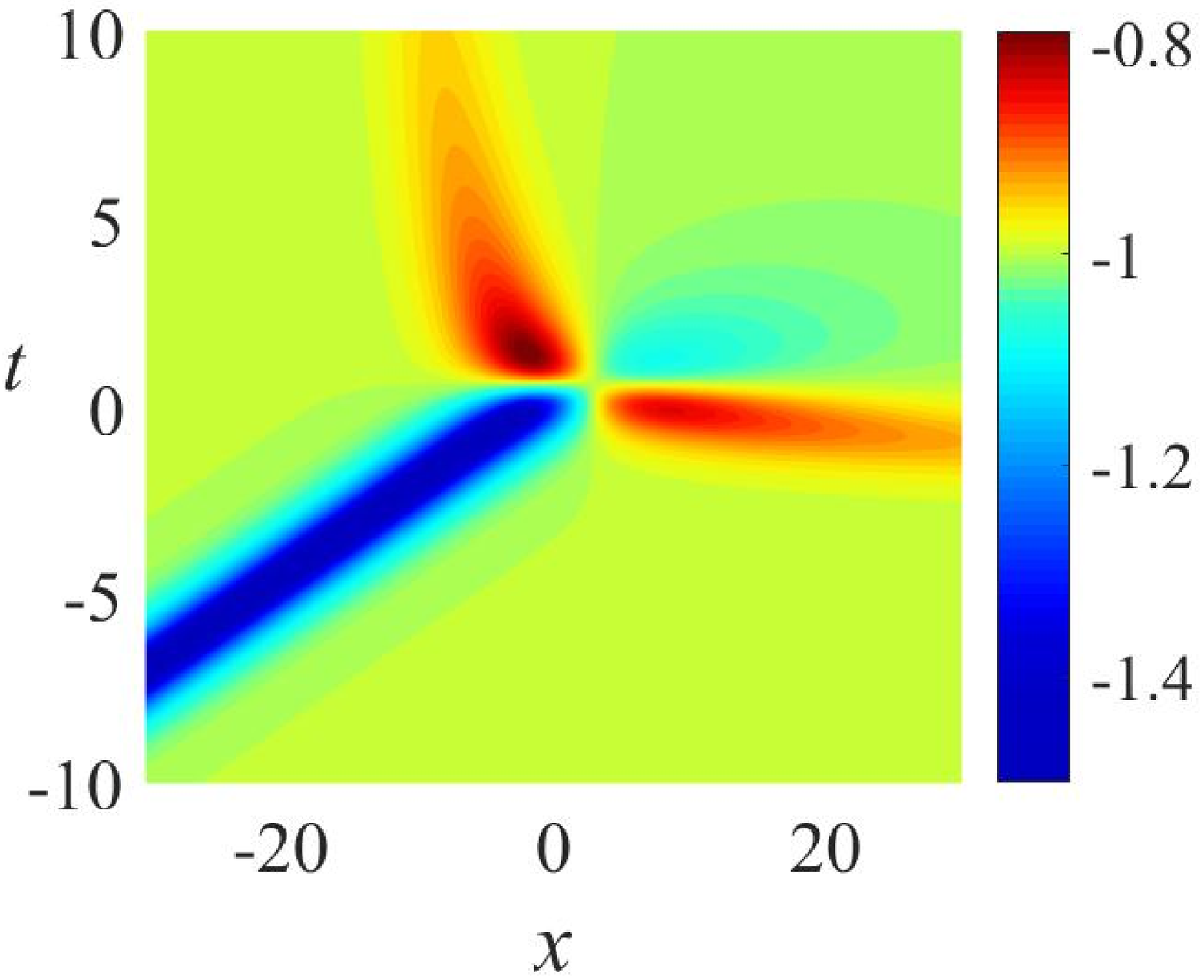}
\end{minipage}
}
\subfigure[]{
\begin{minipage}[t]{0.33\linewidth}
\centering
\includegraphics[width=4.75cm]{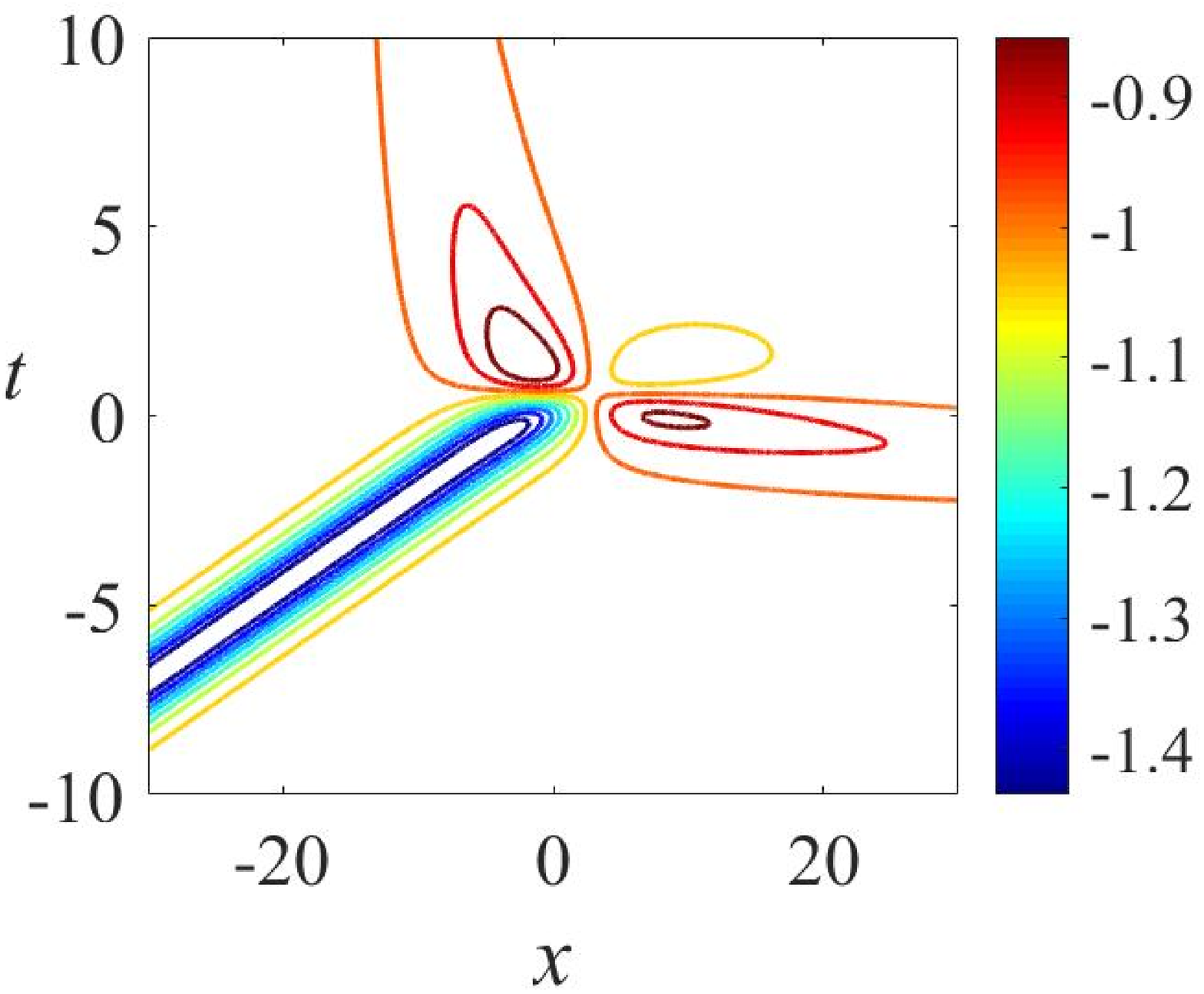}
\end{minipage}%
}

\subfigure[]{
\begin{minipage}[t]{0.33\linewidth}
\centering
\includegraphics[width=4.5cm]{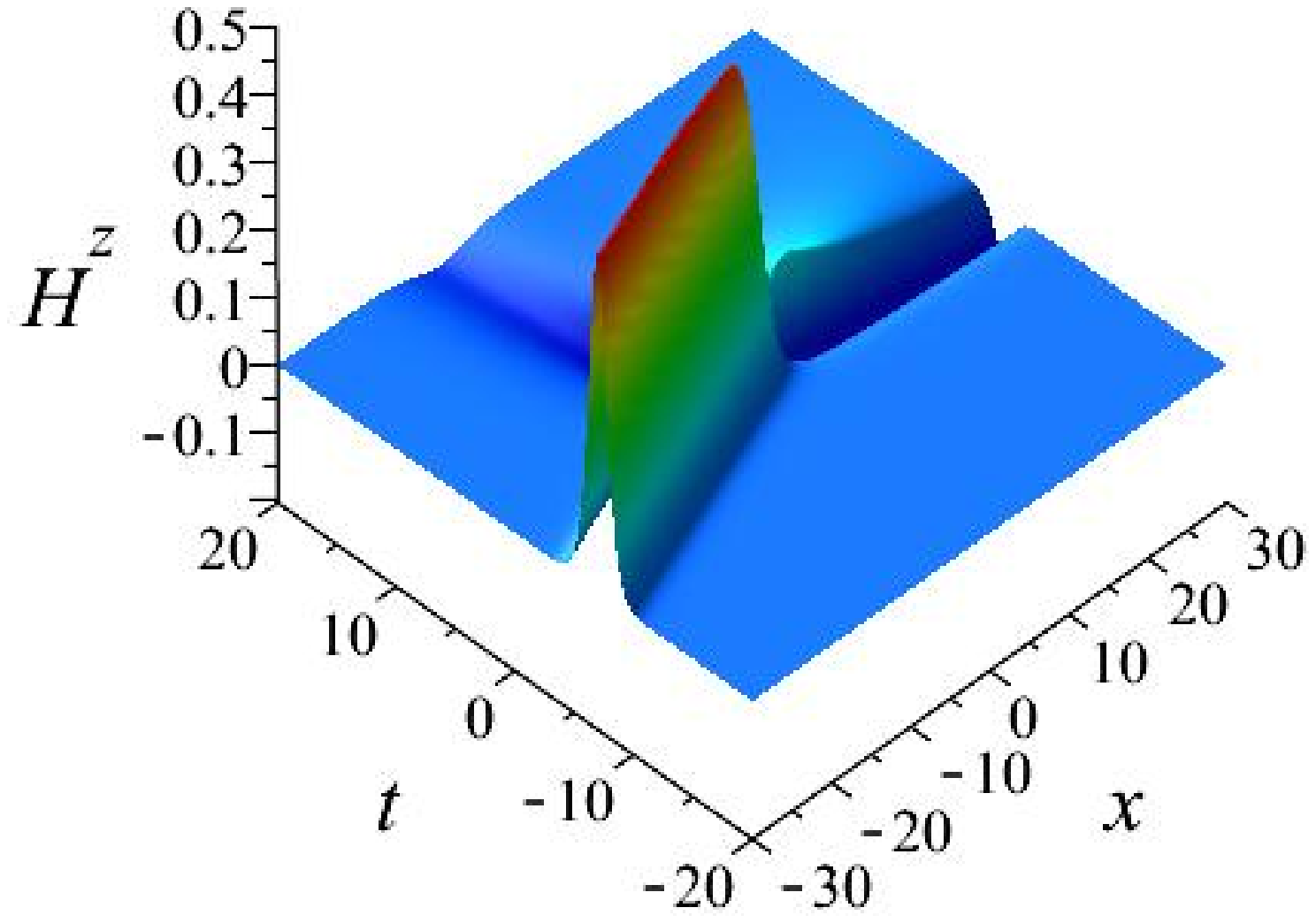}
\end{minipage}
}
\subfigure[]{
\begin{minipage}[t]{0.33\linewidth}
\centering
\includegraphics[width=4.75cm]{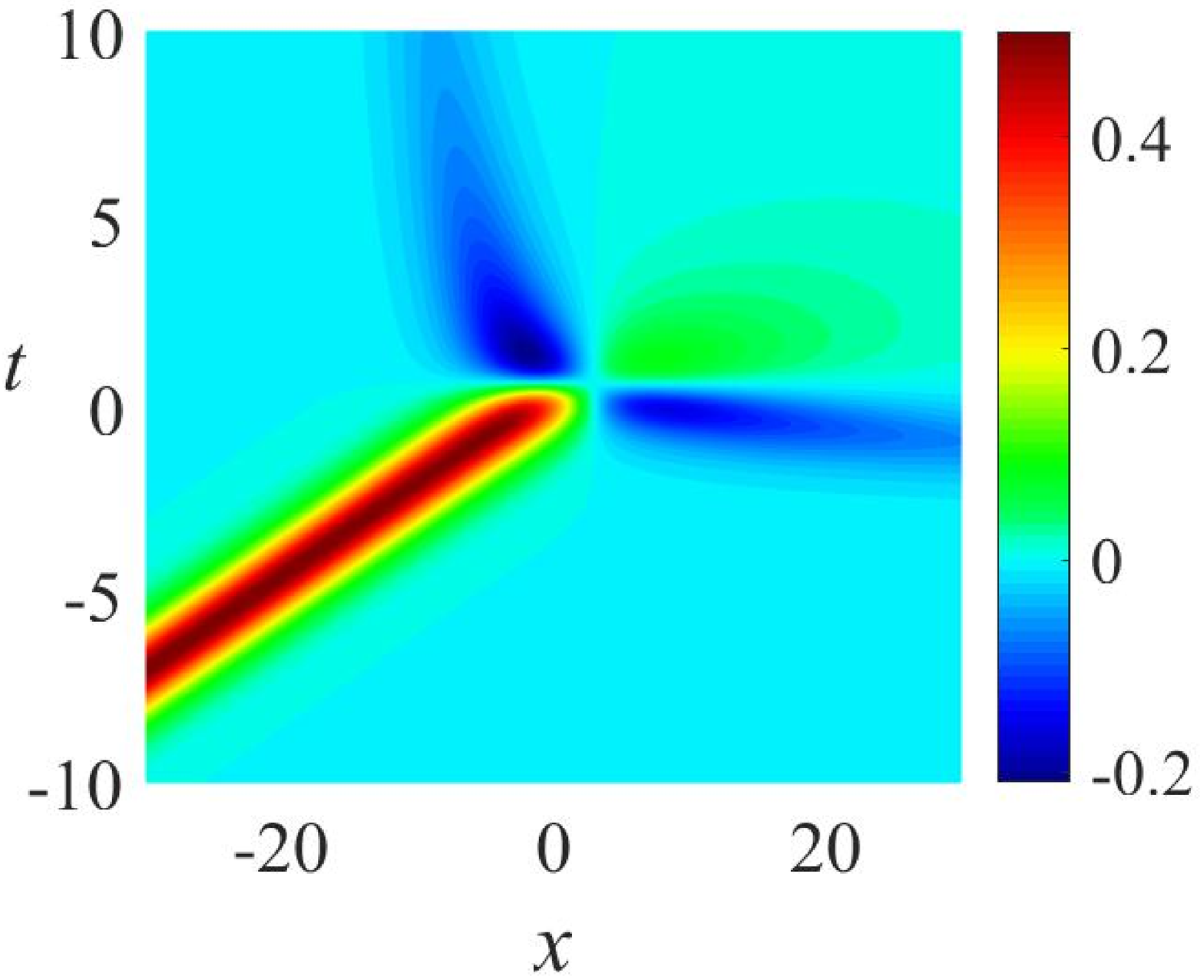}
\end{minipage}%
}
\subfigure[]{
\begin{minipage}[t]{0.33\linewidth}
\centering
\includegraphics[width=4.75cm]{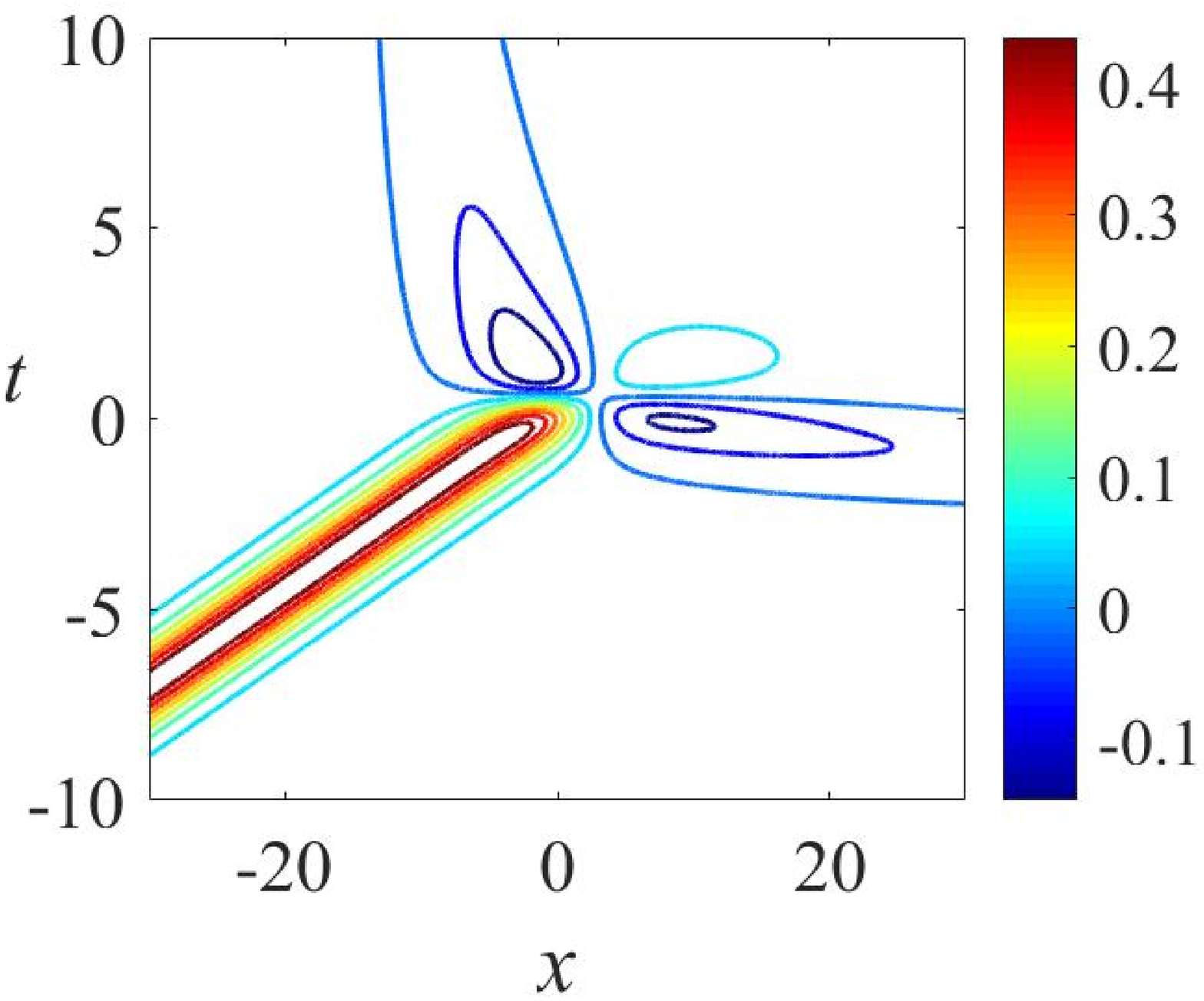}
\end{minipage}
}
\caption{The evolution plots to the Y-type interaction solution (30) of $H^{y}$ and $H^{z}$ with $k_{1}=-1/6,l_{1}=1,k_{2}=1/3,\omega_{2}=1,l_{2}=1/2,a_{0}=1,\omega_{3}=-2,l_{3}=3,k_{4}=-1/2,l_{4}=2$: (a) evolution plot of $H^{y}$, (b) density plot of $H^{y}$, (c) contour plot of $H^{y}$, (d) evolution plot plot of $H^{z}$, (e) density plot of $H^{z}$, (f) contour plot of $H^{z}$.}
\end{figure}
Fig.(7) shows this Y-type interaction solution with parameters selected as $k_{1}=-1/6,l_{1}=1,k_{2}=1/3,\omega_{2}=1,l_{2}=1/2,a_{0}=1,\omega_{3}=-2,l_{3}=3,k_{4}=-1/2,l_{4}=2$. The evolution plots of $H^{y}$ and $H^{z}$ are described in Figs.7(a) and 7(d). Figs.7(b) and 7(e) represent the density plots of the interaction solution. Figs.7(c) and 7(f) show the contour plots of the interaction solution.
\section{Exact solutions for the damped-KMM system}
Now, we investigate the exact solutions for the damped-KMM system ($s\ne0$). As mentioned in section 1, for this non-integrable system, analytical expression of the associated solution has not been provided. We pursue the same transformation as section 2. Finally, the damped-KMM system can be rewritten as follows
\begin{equation}
\begin{split}
&u_{t}=vv_{x},\\
&v_{xt}=uv-sv_{x}.
\end{split}
\end{equation}
Similarly, for Eq.(31), we can take the truncated tanh function expansions as
\begin{equation}
\begin{split}
u=u_{0}+u_{1}\tanh(f)+u_{2}\tanh^{2}(f),\ \ \ \ v=v_{0}+v_{1}\tanh(f).
\end{split}
\end{equation}
By calculating similarly to the section 2, we derive the solution of the damped-KMM system (31)
\begin{equation}
\begin{split}
u=2f_{x}f_{t}\tanh^{2}(f)-(2f_{xt}+\frac{2sf_{x}}{3})&\tanh(f)-\frac{f_{x}s^{2}}{9f_{t}}+\frac{(2f_{xt}f_{t}-f_{x}f_{tt})s}{3 f_{t}^{2}}+\frac{f_{xtt}f_{t}-f_{xt}f_{tt}}{f_{t}^{2}}-2f_{x}f_{t},\\
&v=2f_{t}\tanh(f)-\frac{f_{tt}}{f_{t}}-\frac{s}{3},
\end{split}
\end{equation}
with $f$ satisfied the following tri-linear equations
\begin{equation}
\begin{split}
&f_{x}f_{t}^{2}s^{3}-(6f_{xt}f_{t}^{2}-6f_{x}f_{t}f_{tt})s^{2}-(36f_{x}f_{t}^{4}-9f_{x}f_{tt}^{2}-18f_{xtt}f_{t}^{2}+36f_{xt}f_{t}f_{tt})s\\
&-108f_{xt}f_{t}^{4}+27f_{xttt}f_{t}^{2}-27f_{xt}f_{t}f_{ttt}-81f_{xtt}f_{t}f_{tt}+81f_{xt}f_{tt}^{2}=0,\\
\end{split}
\end{equation}
\begin{equation}
\begin{split}
&(f_{x}f_{t}f_{tt}-f_{xt}f_{t}^{2})s^{2}+(6f_{x}f_{tt}^{2}-3f_{x}f_{t}f_{ttt}+3f_{xtt}f_{t}^{2}-6f_{xt}f_{t}f_{tt})s-36f_{xt}f_{t}^{4}\\
&+9f_{xttt}f_{t}^{2}-9f_{xt}f_{t}f_{ttt}-27f_{xtt}f_{t}f_{tt}+27f_{xt}f_{tt}^{2}=0.
\end{split}
\end{equation}
A simple solution of Eqs.(34) and (35) has the form
\begin{equation}
\begin{split}
f=k_{0}x+\omega_{0}t,\ \ \ \ \omega_{0}=\pm\frac{s}{6},
\end{split}
\end{equation}
where $k_{0}$ is a free constant. The result (we select the symbol "$\pm$" as "$+$" without loss of generality) yields a single soliton solution
\begin{equation}
\begin{split}
&u=\frac{ks}{3}\left(\tanh^{2}(kx+\frac{st}{6})-2\tanh(kx+\frac{st}{6})-3\right),\\
&v=\frac{s}{3}\left(\tanh(kx+\frac{st}{6})-1\right).
\end{split}
\end{equation}
This implies that, for a given damped ferrite material, only soliton with a specific velocity can propagate steadily. Moreover, in order to balance the damping effect, the expression of $u_{1}$ in (32) cannot be eliminated, which makes the propagation mode of $H^{y}$ changes from soliton to kink solution.

To seek interaction solution of Eq.(31), the function $f$ is expressed as
\begin{equation}
\begin{split}
f=k_{0}x+\omega_{0}t+F(k_{1}x+\omega_{1}t).
\end{split}
\end{equation}
By the CTE method, it is not difficult to verify that Eqs.(34) and (35) possess the following solution
\begin{equation}
\begin{split}
&u=\frac{(k+Q_{x})s}{3}\left(\tanh^{2}(kx+\frac{st}{6}+Q)-2\tanh(kx+\frac{st}{6}+Q)-3\right),\\
&v=\frac{s}{3}\left(\tanh(kx+\frac{s}{6}+Q)-1\right),
\end{split}
\end{equation}
where $Q$ is an arbitrary function of $x$. It is actually the existence of such an arbitrary function which provides the way to generate a set of various non-trivial solutions. For instance, when selecting $Q=\sech(x)\cos(2x)$, a special interaction solution is depicted in Fig.8. As we can see in this figure, the few-cycle pulse of component $H^{y}$ moves to the positive half-axis of $t$-axis and then collides inelastically with a kink soliton. At the interaction area, the few-cycle pulse decays quickly and ultimately absorbed by the kink soliton. Simultaneously, the bright soliton of the component $H^{z}$ oscillates up and down and excites an observable wave packet. We actually believe that the above analytical solutions to the damped-KMM equation would arguably offer a matter of investigation to generate diverse propagation modes in real ferrites.
\begin{figure}[htbp]
%
\subfigure[]{
\begin{minipage}[t]{0.5\linewidth}
\centering
\includegraphics[width=7cm]{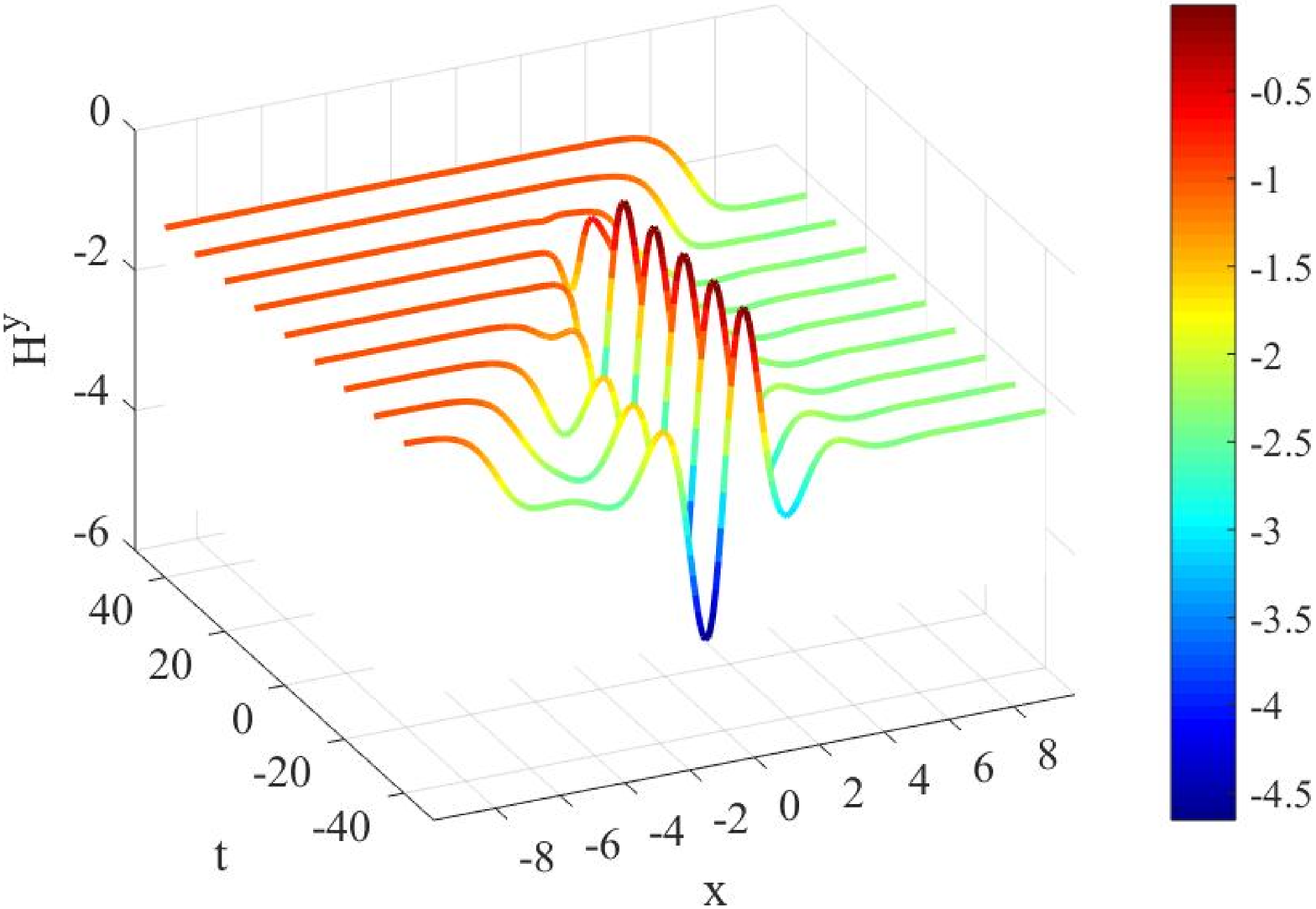}
\end{minipage}%
}
\subfigure[]{
\begin{minipage}[t]{0.5\linewidth}
\centering
\includegraphics[width=7cm]{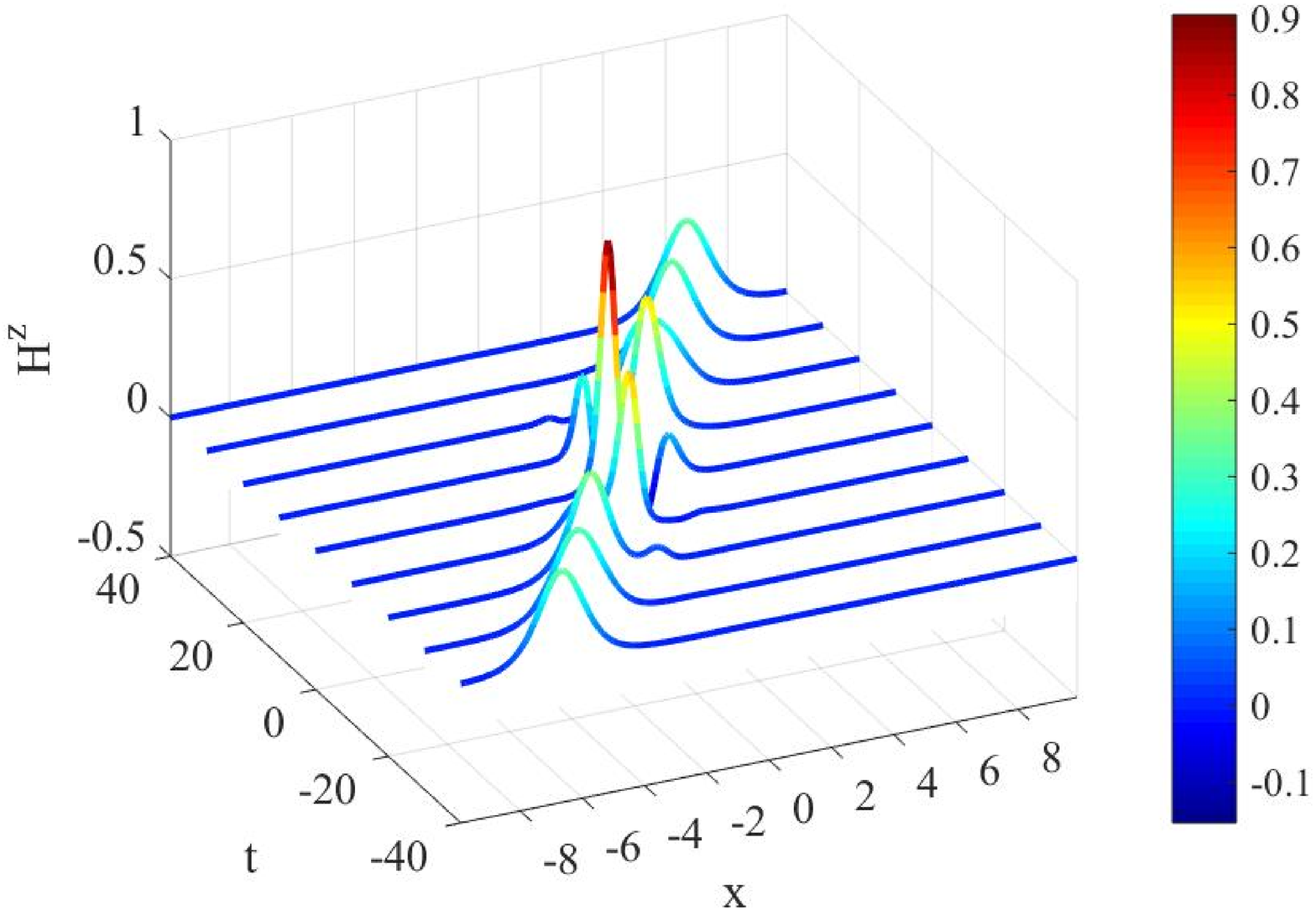}
\end{minipage}
}
\caption{Propagations of interaction solution (39): (a) $H^{y}$ and (b) $H^{z}$ in (39) with $s=-1,k=1$.}
\end{figure}

\section{Summary and discussions}
In summary, we have paid attention to a nonlinear system describing the propagation of short-waves in ferrite, in which Gilbert damping effects have been taken into account. This system, namely the damped-KMM system, has not been provided any analytic solution since it is non-integrable. On the other hand, the damped-KMM system can be reduced to the original KMM system when vanishing the dissipative parameter $s=0$, which has been investigated under many tractable methods. In the present work, the CTE method is applied to the original KMM system, two classes of new interaction solutions are obtained. Particularly, the breather soliton, periodic oscillation soliton and multipole instanton are obtained and shown in Fig.(1)-(3), respectively. Beyond that, one can construct many other kinds of solutions because the solutions include an arbitrary function. Based on the Painlev$\acute{\rm e}$ analysis, a coupled equation including a quadri-linear form and a tri-linear form of the original KMM system is obtained. The rogue wave solution is given by introducing a quadratic function. In addition, the interaction solutions between rogue wave and multi-solitons are proposed by combining the quadratic function with additional exponential functions. What's more, we have constructed solutions of the damped-KMM system at the first time. The analytical expressions (39) as well as Fig.(8) show that amplitude of the wave decreases as time evolves. In fact, in real ferrites with damping effect, the damping involves loss of energy from the macroscopic motion of the local magnetization field by transfer of kinetic and potential energies to microscopic thermal motion in the form of spin waves, lattice vibration, and thermal excitations, among others \cite{34}. In addition, Kuetche and coauthors \cite{42,43} have proposed a generalized KMM system, that takes into account the inhomogeneous exchange along with the Gilbert-damping effect. Further contribution on such inhomogeneous exchange effects are needed to understand more deeply the behavior of the wave in magnetic insulators \cite{44}. We believe that the results of this work would be worth underlying in the future investigations.

\section{Acknowledgments}
This work was supported by the National Natural Science Foundation of China under Great Nos. 11835011 and 11675146.



\end{document}